\documentclass[%
reprint,
superscriptaddress,
nofootinbib,
 amsmath,amssymb,
 aps,
prb,
]{revtex4-2}

\usepackage{graphicx}% Include figure files
\usepackage{dcolumn}% Align table columns on decimal point
\usepackage{bm}% bold math
\usepackage{xcolor}
\begin{document}

\preprint{APS/123-QED}

\title{Creating and detecting poor man’s Majorana bound states
in interacting quantum dots}%

\author{Athanasios Tsintzis}
\affiliation{%
Division of Solid State
Physics and NanoLund, Lund University, S-221 00 Lund, Sweden
}%
\author{Rub\'en Seoane Souto}%
\affiliation{%
Division of Solid State
Physics and NanoLund, Lund University, S-221 00 Lund, Sweden
}%
\affiliation{%
Center for Quantum Devices, Niels Bohr Institute, University of
Copenhagen, DK-2100 Copenhagen, Denmark
}%
\author{Martin Leijnse}
\affiliation{%
Division of Solid State
Physics and NanoLund, Lund University, S-221 00 Lund, Sweden
}%
\affiliation{%
Center for Quantum Devices, Niels Bohr Institute, University of
Copenhagen, DK-2100 Copenhagen, Denmark
}%

\date{\today}% It is always \today, today,
             %  but any date may be explicitly specified

\begin{abstract}
We propose and theoretically investigate an alternative way to create the poor man's Majorana bound states (MBSs) introduced in Phys. Rev. B {\bf 86}, 134528 (2012). Our proposal is based on two quantum dots (QDs) with strong electron-electron interactions that couple via a central QD with proximity-induced superconductivity. In the presence of spin-orbit coupling and a magnetic field, gate control of all three QDs
%Gating the central QD provides control of the relative amplitudes of crossed Andreev reflection and elastic cotunneling between the outer QDs. This control, in the presence of spin-orbit coupling and a magnetic field,
allows tuning the system into sweet spots with one MBS localized on each outer dot. We quantify the quality of these MBSs and show how it depends on the Zeeman energy and interaction strength. We also show how nonlocal transport spectroscopy can be used to identify sweet spots with high MBS quality. Our results provide a path for investigating MBS physics in a setting that is free of many of the doubts and uncertainties that plague other platforms.

\end{abstract}

\maketitle

%\tableofcontents
%%%%%%%%%%%%%%%%%%%%%%%%%%%%%%%%
%%%%%%%% Introduction %%%%%%%%%%
%%%%%%%%%%%%%%%%%%%%%%%%%%%%%%%%

{\it Introduction.}
The realization of Majorana bound states (MBSs)~\cite{Alicea_RPP2012,LeijnseReview,AguadoReview,BeenakkerReview_20} is one of the most heavily pursued goals in condensed matter physics. The motivation is their theoretically predicted nonabelian and nonlocal properties. In addition to being of fundamental interest as a new physics phenomenon, these properties allow for protected ways to store and manipulate quantum information~\cite{NayakReview}. The simplest toy model where MBSs arise is the Kitaev model~\cite{Kitaev_2001}, a tight-binding chain with spinless electrons and $p$-wave superconducting pairing. An explosion in experimental activities was motivated by various theoretical proposals showing that different quantum systems can be engineered such that the Kitaev model arises as an effective description of the low-energy degrees of freedom~\cite{Lutchyn_PRL2010, Oreg_PRL2010, PhysRevB.88.020407, Hell2017, Pientka2017, Vaitiekenas2020, flensberg2021engineered}. 

By now, experiments have revealed signatures consistent with MBSs in several of the proposed platforms, see Refs.~\cite{Mourik_science2012, deng2012anomalous, finck2013anomalous, NadjPerge2014, deng2016majorana, Nichele_PRL2017, lutchyn2018majorana, Fornieri2019, Ren2019, Vaitiekenas2020} for a few examples. However, it has also become increasingly clear that the disorder that plagues all real materials can give rise to other, nontopological, states that can provide an alternative explanation for most experimental observations~\cite{Prada_PRB2012,Kells_PRB12,Liu2012,Liu2017, Moore_PRB18,reeg2018zero,Awoga_PRL2019,Vuik_SciPost19,Pan_PRR20,Prada_review,hess2021local}. So far, the nonabelian and nonlocal properties of MBSs have not been experimentally demonstrated. 

One  way to avoid the problem of imperfect materials is to engineer an artificial Kitaev chain in quantum dots (QDs) coupled via narrow superconducting regions~\cite{Sau_NatComm2012}. In fact, it was shown in Ref.~\cite{Leijnse_PRB2012} that two QDs are enough to obtain MBSs, named poor man's MBSs because they possess all the properties of MBSs but only exist at fine-tuned sweet spots in parameter space. The poor man's MBSs system closely resembles Cooper pair splitter devices~\cite{Recher_PRB2001,Hofstetter_Nature2009,Herrmann_PRL2010,Fulop_PRL2015},
%,Burset_PRB2011,Das_NatComm2012,Schindele_PRL2012,Braunecker_PRL2013,Tan_PRL2015,Fulop_PRL2015,Dominguez_PhysE2016,Borzenets_SRep2016}
but requires both strong crossed Andreev reflection (CAR) and the ability to fine-tune either the spin-orbit coupling strength or the angle between non-collinearly polarized QD spins. In addition, given the lack of topological protection, it is unclear how close one can come to ideal MBSs in a realistic system with finite Zeeman energy and electron-electron interactions on the QDs.  
 
In this work, we show a way to overcome the difficulties and uncertainties associated with the original proposal for poor man's MBSs. A key ingredient is to couple the QDs via a central QD which is, in turn, strongly proximitized by a superconductor. The advantage is that gating the center QD provides control of the relative amplitudes of CAR and elastic cotunneling (ECT), which allows realizing poor man's MBSs with a constant spin-orbit coupling (or a constant finite angle between the effective magnetic fields on the two QDs). The underlying physics is the same as a recent proposal for coupling the QDs via an Andreev bound state~\cite{Liu_arXiv2022}.
We analyze the role of finite Zeeman splitting (including both spin states on all three QDs) as well as strong Coulomb interactions. We show that sweet spots in parameter space exist where the system exhibits three characteristics that are prerequisites for MBSs with nonabelian properties: $(i)$ degenerate even and odd (electron number) parity ground states; $(ii)$ a substantial gap to the excited states; $(iii)$ localized MBSs of high quality, which we quantify with the Majorana polarization (MP)~\cite{Aksenov2020,Sedlmayr2015,Sedlmayr2016}. Our results also show how the MP depends on the interaction strength and Zeeman energy. This is important because there are regions in parameter space associated with apparent sweet spots that fulfill $(i)$ and $(ii)$, but have poor MP. Finally, we calculate the nonlocal transport signatures of the interacting system, and show that they can be used to identify sweet spots and distinguish between true sweet spots and apparent sweet spots with low MP. 

While finalizing the present manuscript, a report of experimental signatures consistent with poor man's MBSs appeared~\cite{Dvir_arXiv2022} based on the Andreev bound state coupling proposed in Ref.~\cite{Liu_arXiv2022}. 
%It is based on coupling QDs via an Andreev bound state that can be used to tune to the MBS sweet spot~\cite{Liu_arXiv2022} in a way that is similar to our proposal. 
The experiments were compared with a non-interacting model with infinite Zeeman energy, similar to Ref.~\cite{Leijnse_PRB2012}.
%which does not account for excited spin states and electron-electron interactions, and does not reveal the quality of the MBSs. 
The ground-state properties of interacting double QDs harbouring poor man's MBSs have also been investigated in Refs.~\cite{PhysRevB.88.144515,Wright_PRL2013, OBrien2015}.
%\begin{itemize}
%    \item General - Majorana - Kitaev \cite{Kitaev2001},  %topological quantum computation
%    \item Early theoretical models \cite{Lutchyn2010,Oreg2010} %and experiments (Refs?) (Nanowires). Full shell as well? 
%problems with regular approach    
%    \item The poor man's approach, QD-S well established and studied + CP splitters \cite{poormans_original}
%    \item problems with poormans
%    \item our system
%    \item Mention Delft papers? \cite{Liu2022,Wang2022,Dvir2022}
%\end{itemize}

%Poor man's Majorana bound states \cite{poormans_original}. Majorana polarization (MP) for interacting systems \cite{MP_mb}. Local pairing $\Delta$

\begin{figure}
	\includegraphics[width=0.42\textwidth,trim={0cm 0cm 0cm 0cm},clip]{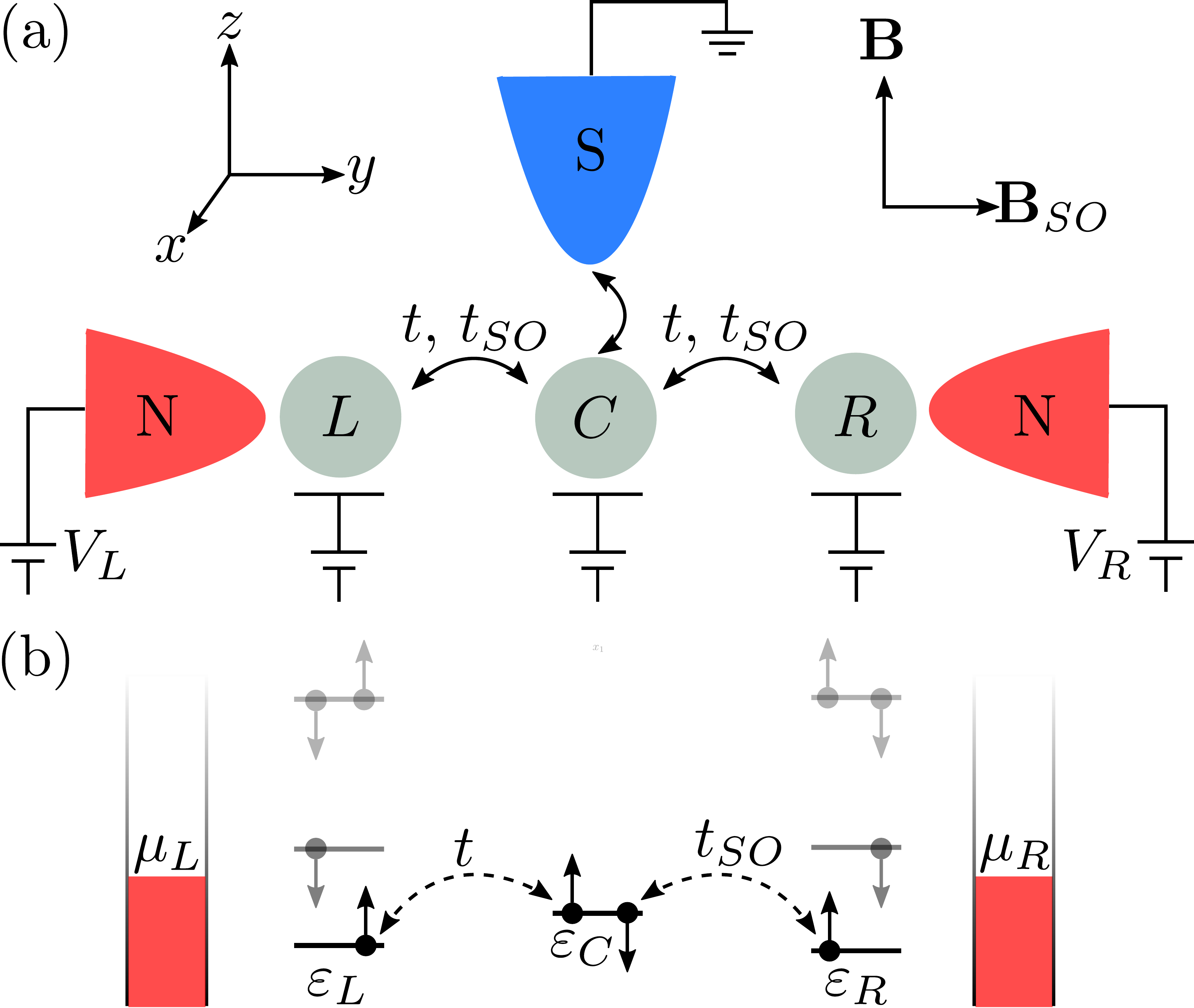}
	\caption{(a)~Setup with three QDs ($L,C,R$) coupled via spin-conserving tunneling $t$ and (spin-orbit induced) spin-flip tunneling $t_{SO}$. QD $C$ is strongly coupled to a grounded superconductor. $\mathbf{B}$ is an external Zeeman field and $\mathbf{B}_{SO}$ is a spin-orbit field. (b)~QD orbitals $\varepsilon_j$ and examples of tunnel processes. Electron-electron interactions increase the energy cost of occupying a QD with two electrons. Two normal leads with chemical potentials $\mu_{L,R}$ (controlled by voltages $V_{L,R}$) couple to QDs $L$ and $R$ and can be used for tunnel spectroscopy.}
	\label{fig1}
\end{figure}
%\section{System}

%%%%%%%%%%%%%%%%%%%%%%%%%%%%%%%%%%%%%%
%%%%%%%%%%%%%%% Model %%%%%%%%%%%%%%%%
%%%%%%%%%%%%%%%%%%%%%%%%%%%%%%%%%%%%%%

{\it Proposed device and model.}
To create and detect poor man's MBSs, we propose a device with three coupled QDs. The setup is shown in Fig.~\ref{fig1}(a), while Fig.~\ref{fig1}(b) shows a sketch of the involved energies and tunnel processes. The system is described by the Hamiltonian [excluding for now the normal (N) leads]:
%\begin{equation}
%H = H_{QD} + H_{\mathbf{B}} + H_{SO} + %H_{\Delta},
%\end{equation}
%where
\begin{eqnarray}
H_{QDs} &&= \sum_{\sigma, j} \varepsilon_j n_{j \sigma} + \sum_j U_j n_{j \uparrow} n_{j \downarrow} + \sum_j E_{Zj} n_{j \downarrow} \nonumber \\
&&+ \sum_{\sigma, j \neq C} \Big[ t_{j} d_{j \sigma}^\dagger d_{C \sigma} + h.c. \Big]  \nonumber \\
&&+ \sum_{j\neq C} \Big[ t^{SO}_{j} d_{j \uparrow}^\dagger d_{C \downarrow} - t^{SO}_{j} d_{j \downarrow}^\dagger d_{C \uparrow} + h.c.\Big]\nonumber \\
&&+\Delta \Big[ d_{C \uparrow}^\dagger d_{C \downarrow}^\dagger + h.c. \Big].\label{eq:HQD}
\end{eqnarray}
%ML I changed the t and t_SO terms because the old versions would contain a hopping between 1 and 3 (only restriction j<k)
%and
%begin{equation}
%H_{\Delta} = \Delta \Big[ d_{C \uparrow}^\dagger %d_{C \downarrow}^\dagger + H.C. \Big].
%\end{equation}
Here, $d_{j\sigma}^\dagger$ creates an electron with spin $\sigma=\uparrow, \downarrow$ in QD $j=L,C,R$ with occupation $n_{j \sigma} = d_{j \sigma}^\dagger d_{j \sigma}$, single particle orbital energy $\varepsilon_j$, charging energy $U_j$ and Zeeman energy $E_{Zj}$. $t_{j}$ is the amplitude for spin-conserving tunneling between QDs $j=L,R$ and QD $C$, while $t_j^{SO}$ is the amplitude for spin-flip tunneling which results from a spin-orbit interaction with spin-orbit field $\mathbf{B}_{SO}$ along the $y$-axis, perpendicular to the external Zeeman field $\mathbf{B}$ chosen here to be along the $z$-axis~\cite{Stepanenko2012}. We include the proximity-induced superconductivity on QD $C$ through a pairing term of amplitude $\Delta$, which is a reasonable approximation for energies below the superconducting gap~\cite{Governale2008,Rozhkov2000,Tanaka2007,Karrasch2008}.
%For $H_{SO}$ we use the expression (cf. Ref. \cite{Stepanenko2012}):
%\begin{equation}
%H_{SO} = \mathbf{B}_{SO} \cdot \sum_{\sigma, \sigma'} \Big[ i d_{j \sigma}^\dagger \mathbf{S}^{\sigma \sigma'} d_{k \sigma'} + H.C.\Big],
%\end{equation}
%where $\mathbf{B}_{SO} = t_{SO} \, \hat{\mathbf{y}}$ is the spin-orbit field and $\mathbf{S}$ is the Pauli vector.

{\it Relation to the original poor man's MBS model.} 
In the original model for poor man's MBSs~\cite{Leijnse_PRB2012}, a superconductor mediates two different types of couplings between two fully spin-polarized QDs: CAR and ECT -- corresponding respectively to the pairing and hopping terms in the Kitaev model~\cite{Kitaev_2001}. The CAR and ECT amplitudes scale in the same way with the QD-superconductor coupling strength, but their ratio can be controlled via the angle between the QD spins. The MBS sweet spot occurs when the CAR and ECT amplitudes are equal and both QD levels are at zero energy (i.e., aligned with the chemical potential of the superconductor). At this point, the Kitaev chain hosts one MBS fully localized on each end site; then two sites -- or two QDs -- suffice to have spatially separated MBSs. 

The relation between $H_{QDs}$ in Eq.~(\ref{eq:HQD}) and the simple poor man's MBS model is most easily understood in the regime where $|\varepsilon_j|, |t_j|, |t_j^{SO}|, |E_{ZC}| \ll |\Delta|, |E_{ZL,R}|$ (although our future analyses will not be limited to this regime). Then, in the ground state, QDs $L$ and $R$ are occupied by zero or one electron each, while QD $C$ is in a superposition of empty and doubly occupied (single occupation being suppressed by the large superconducting pairing). Second order perturbation theory in $t_j$ and $t_j^{SO}$ gives a coupling between QDs $L$ and $R$, both through ECT ($\propto t_L t_R$) and CAR ($\propto t_L t_R^{SO} + t_R t_L^{SO}$, as the singlet nature of the Cooper pairs means that a spin flip is needed to populate the lowest spin state of each QD). $H_{QDs}$ conserves the parity (even or odd) of the total electron number. Couplings within the even (odd) parity sector are mediated by CAR (ECT) which therefore lowers the energy of the even (odd) parity ground state. However, because of interference between different tunnel processes, the amplitudes of CAR and ECT depend differently on $\varepsilon_C$, such that ECT is suppressed around $\varepsilon_C = 0$. A similar control of the CAR and ECT relative amplitudes can be achieved using a closely related model with an Andreev bound state
%an ABS 
mediating the coupling between two QDs, as proposed in Ref.~\cite{Liu_arXiv2022} and exploited in the experiments presented in Refs~\cite{Dvir_arXiv2022, Wang2022}. 

Based on the original poor man's MBS model, a sweet spot is expected when the ECT and CAR amplitudes are  equal and $\varepsilon_L = \varepsilon_R = 0$. To some degree this still holds in our model for finite $E_{Zj}$ and $U_j$, but we need to compensate for renormalizations of $\varepsilon_{L,R}$ due to the coupling to QD $C$. Away from the perturbative regime (in $t_j, t_j^{SO}$), the concepts of CAR and ECT are no longer well-defined, but the processes coupling states within the even parity sector and within the odd parity sector still depend differently on $\varepsilon_C$. % Furthermore, the detailed calculations presented below are needed to address several crucial questions: Is there still a sweet spot for the interacting system with realistic parameter values? If so, what are the requirements for reaching it? Will the MBSs be well-localized, have the same robustness to variations in parameters as in the simpler model, and have a large enough gap to excited states? Given that interactions often significantly change the transport properties, can we detect the sweet spot in the same way as in the simple non-interacting model, and can we use transport spectroscopy to gain information about the quality of the MBSs at the sweet spot? 

%%%%%%%%%%%%%%%%%%%%%%%%%%%%%%%%%%%%%%
%%%%%%%%%%%%%%% Results %%%%%%%%%%%%%%%%
%%%%%%%%%%%%%%%%%%%%%%%%%%%%%%%%%%%%%%
{\it Sweet spots and MBS quality.}
Throughout the rest of the paper we, for simplicity, consider a symmetric system, $t_L=t_R=t$,  $t^{SO}_L=t^{SO}_R=t_{SO}$, $U_L=U_R=U$, $E_{ZL}=E_{ZR}=E_Z$. We assume that the strong coupling of QD $C$ to the grounded superconductor has quenched its charging energy by a combination of capacitive effects and tunnel-induced renormalization, and reduced its Zeeman splitting (because of the small g-factor of the superconductor). We therefore take $U_C = E_C = 0$ but have verified that relaxing these assumptions does not qualitatively change the results, see SI~\cite{SI}. Unless otherwise stated, we choose the following values for the remaining parameters ($e=\hbar=k_B=1$): $U=5 \Delta$, $t = 0.5 \Delta$, $t_{SO}=0.2 t, E_Z = 1.5 \Delta$.

\begin{figure}
	\includegraphics[width=0.5\textwidth,trim={0cm 0.9cm 0.02cm 0.cm},clip]{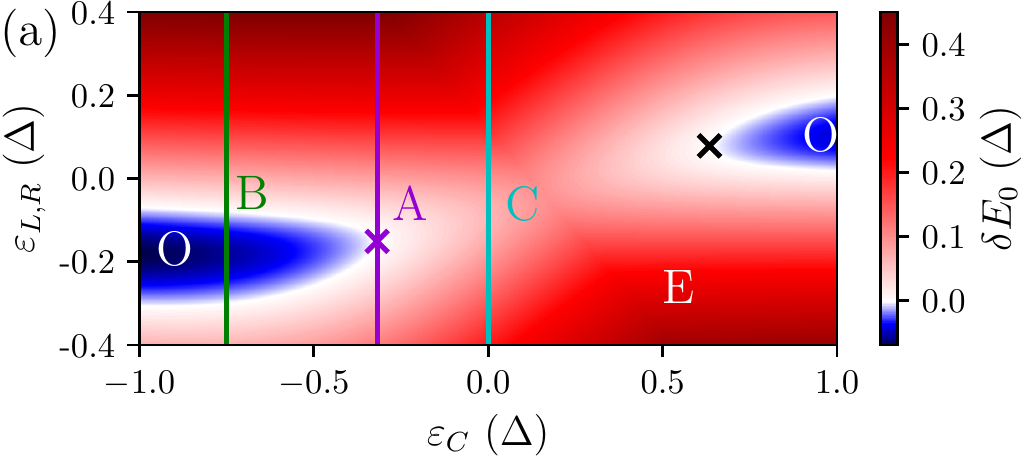}
	\includegraphics[width=0.5\textwidth,trim={0.0cm 0cm -0.0cm 0cm},clip]{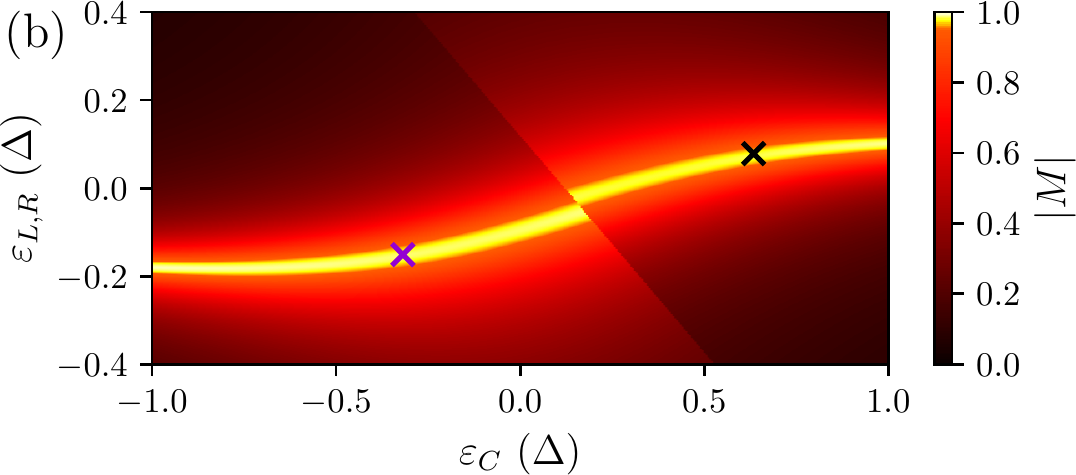}
	\includegraphics[width=0.21\textwidth,trim={0cm 0cm 0cm 0cm},clip]{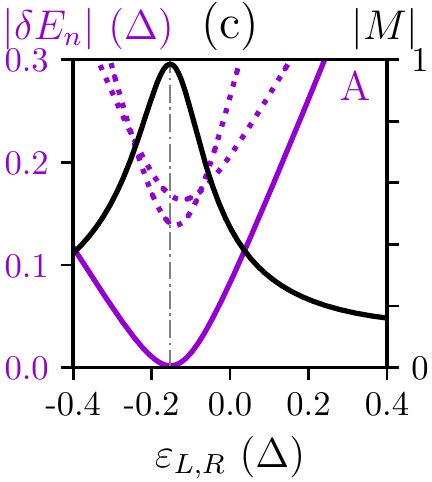}
	\includegraphics[width=0.21\textwidth,trim={0cm 0cm 0cm -0.2cm},clip]{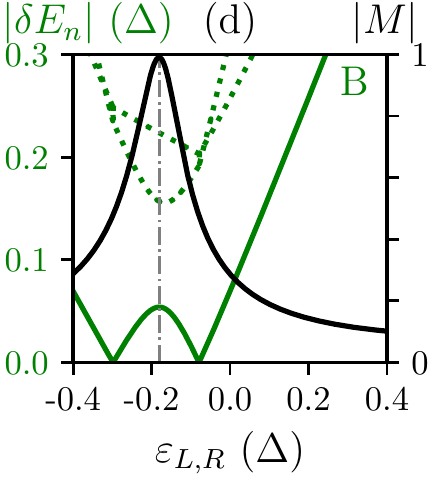}
	\caption{ (a)~$\delta E_0$ as a function of $\varepsilon_L = \varepsilon_R$ and $\varepsilon_C$. Cuts marked A, B and C are explored in (c, d) and in Fig.~\ref{fig_3}, while O and E mark regions where the global ground state has odd and even parity, respectively. The purple cross marks the $\varepsilon_C < 0$ sweet spot at $\varepsilon_C \approx -0.319 \Delta$, $\varepsilon_L = \varepsilon_R \approx -0.151 \Delta$, while the $\varepsilon_C > 0$ sweet spot is marked by the black cross at $\varepsilon_C \approx 0.634 \Delta$, $\varepsilon_L = \varepsilon_R \approx 0.0785 \Delta$.  (b)~$|M|$ as a function of $\varepsilon_L = \varepsilon_R$ and $\varepsilon_C$. The crosses mark the same points as in (a). (c)~$|\delta E_n|$ (left axis, purple lines) as a function of $\varepsilon_L = \varepsilon_R$ along cut A (purple line) in (a). The lowest excited state (full line) has different parity (odd in this case) than the ground state (even in this case); one of the two higher excited states shown (dotted lines) has even parity and the other odd. The black line shows $|M|$ (right axis) as a function of $\varepsilon_L = \varepsilon_R$ along cut A in (a). (d)~Same as (c) but along cut B (green line) in (a). The vertical dash-dotted lines in (c) and (d) indicate the maximum of $|M|$.}
	\label{fig_22}
\end{figure}

\begin{figure*}[htp!]
	\includegraphics[width=0.32\textwidth,trim={0cm 0cm 0cm 0cm},clip]{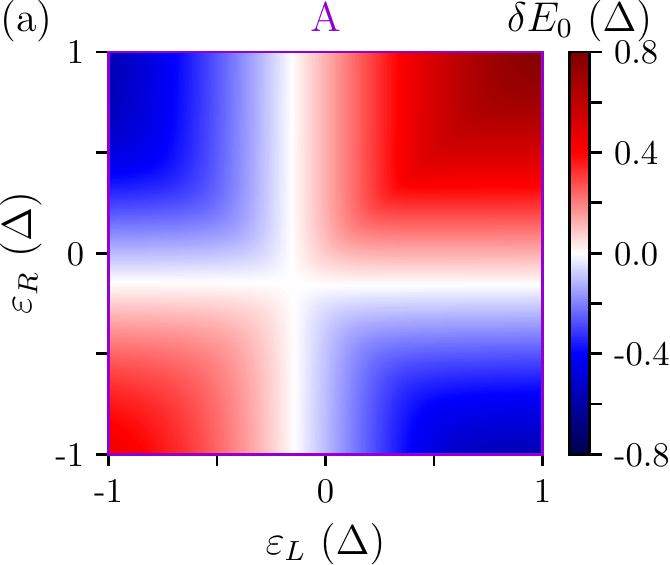}
	\includegraphics[width=0.32\textwidth,trim={0cm 0cm 0cm 0cm},clip]{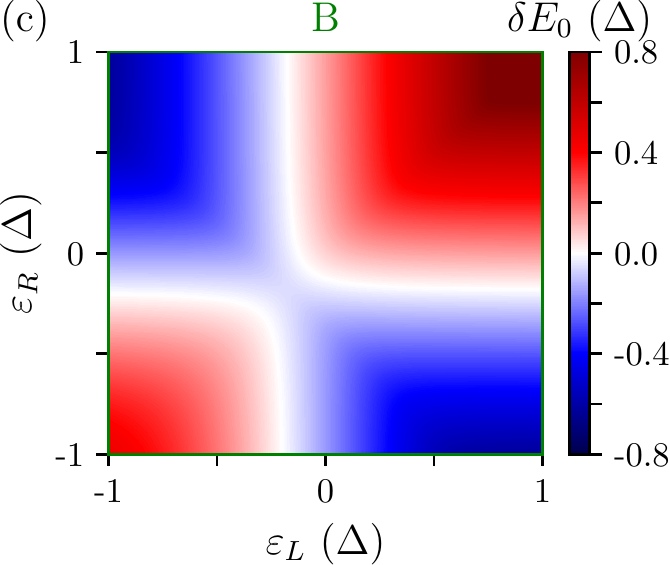}
	\includegraphics[width=0.32\textwidth,trim={0cm 0cm 0cm 0cm},clip]{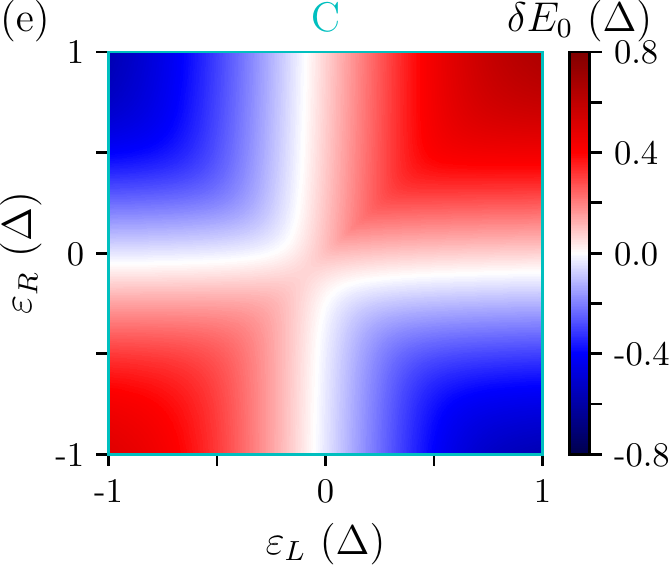}
	\includegraphics[width=0.32\textwidth,trim={0cm 0cm 0cm 0cm},clip]{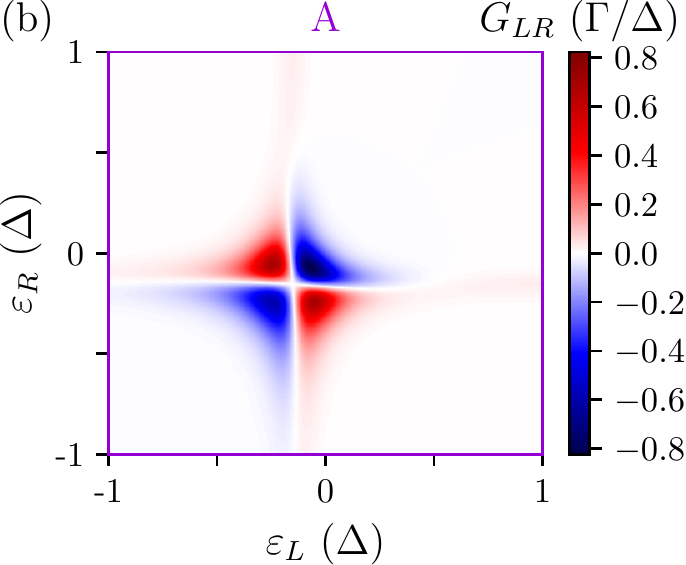}
	\includegraphics[width=0.32\textwidth,trim={0cm 0cm 0cm 0cm},clip]{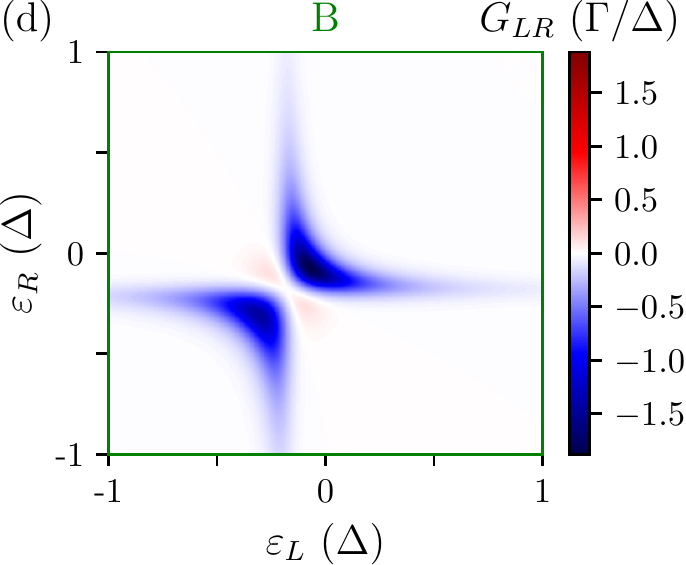}
	\includegraphics[width=0.32\textwidth,trim={0cm 0cm 0cm 0cm},clip]{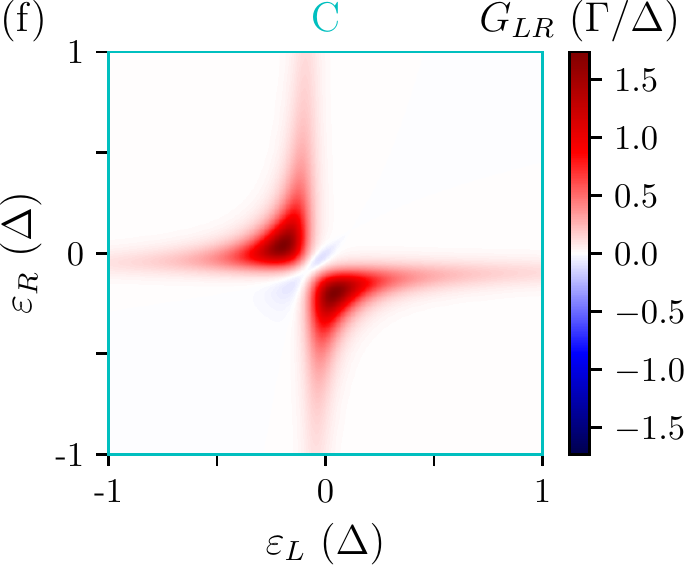}
	\caption{ (a), (c) and (e) Energy difference $\delta E_0$ between even and odd parity ground states as a function of $\varepsilon_L$ and $\varepsilon_R$, with the same $\varepsilon_C$ as in the lines marked A (a), B (c) and C (e) in Fig.~\ref{fig_22}(a). (b), (d) and (f) Same as (a), (c) and (e), but showing $G_{LR}$ instead.}
	\label{fig_3}
\end{figure*}

We first focus on $\delta E_0 = E^O-E^E$, the energy difference between the odd and even parity ground states of $H_{QDs}$ in Eq.~(\ref{eq:HQD}). As explained above, $\varepsilon_C$ affects the even and odd ground state energies by changing the relative strengths of couplings within the even and odd parity sectors. For $|\varepsilon_C| < |\Delta|$ and $\varepsilon_L = \varepsilon_R$, the ground state is dominated by an even electron number on QD $C$. For large positive $\varepsilon_{L,R}$, QDs $L$ and $R$ are both mainly empty, and the global ground state of all three QDs thus has even parity. For negative $\varepsilon_{L,R}$ with $|\varepsilon_{L,R}| < U$, the global ground state is also even, because it is dominated by single occupations of both QDs $L$ and $R$. 

The above behavior is seen in Fig.~\ref{fig_22}(a) which shows $\delta E_0$ as a function of $\varepsilon_C$ and $\varepsilon_L = \varepsilon_R$. When changing $\varepsilon_{L,R}$ along a vertical cut we start and end with an even parity ground state, but if $|\varepsilon_C|$ is large enough  there is a region in between with an odd parity ground state [blue color in Fig.~\ref{fig_22}(a)]. This happens for values of $\varepsilon_C$ such that couplings within the odd parity sector are stronger than those within the even parity sector. There are two values of $\varepsilon_C$ where this region reduces to a point as a function of $\varepsilon_{L,R}$ [marked with purple and black crosses in Fig.~\ref{fig_22}(a)] and we will see that these points are the closest we come to sweet spots with MBSs.

From Fig.~\ref{fig_22}(a) we see that we have lines in parameter space with degenerate even and odd parity ground states (white color). To determine whether these degeneracies are associated with MBSs, we quantify the MBS quality using the MP~\cite{Aksenov2020}. In our case, it corresponds to the degree that a Hermitian operator localized on one of the outer QDs $j\neq C$ can switch between the lowest energy even and odd states:
\begin{eqnarray}
M_j &=& \frac{ \sum_\sigma \left( w_\sigma^2 - z_\sigma^2 \right)}{\sum_\sigma \left( w_\sigma^2 + z_\sigma^2 \right)}, \\
w_\sigma &=& \langle O | (d_{j\sigma} + d_{j\sigma}^\dagger) |E\rangle, \\
z_\sigma &=& \langle O | (d_{j\sigma} - d_{j\sigma}^\dagger) |E\rangle,
\end{eqnarray}
where $|E\rangle$ ($|O\rangle$) is the lowest energy even (odd) parity state. This definition guarantees that $0 \leq |M_j| \leq 1$, where $|M_j| = 1$ would indicate a single MBS perfectly localized on QD $j$, with no other MBS operator having any weight there. For the presented results, $M_L=-M_R$ and in the following we drop the index $j$ and focus on $|M_L| = |M_R| = |M|$ 

Figure~\ref{fig_22}(b) shows $|M|$ plotted over the same range in $\varepsilon_C$ and $\varepsilon_L = \varepsilon_R$ used in Fig.~\ref{fig_22}(a). There is a line where $|M|$ comes very close to 1, but this line only coincides with an even-odd degeneracy at two isolated points (marked with purple and black crosses). The discontinuity in $|M|$ arises because the two lowest energy odd-parity states undergo a crossing. This turns into an avoided crossing for $\varepsilon_L \neq \varepsilon_R$ and occurs in a regime far from even/odd ground state degeneracy where the MP has little meaning.

Figures~\ref{fig_22}(c) and (d) show $|M|$ together with $|\delta E_0|$ and the excitation energies above the global ground state ($|\delta E_n|, n \geq 1$) as a function of $\varepsilon_L = \varepsilon_R$ for two different values of $\varepsilon_C$ [purple and green cuts in Fig.~\ref{fig_22}(a)]. Along both cuts, we find even/odd degeneracies with a substantial separation to excited states. However, it is only in Fig.~\ref{fig_22}(c) that this degeneracy coincides with a large $|M|$ ($\approx 0.986$), while in Fig.~\ref{fig_22}(d) the peak in $|M|$ lies in between the two degeneracy points. Similar results are found for the region where $\varepsilon_C > 0$. 

We also investigate another important property of the MBSs in our system, namely their chargeless nature. For that purpose, we calculate 
\begin{equation}
\delta Q_{L,R} = \langle O| n_{L,R} |O \rangle - \langle E| n_{L,R} |E \rangle,\label{deltaQ}
\end{equation}
i.e., the lowest energy even and odd parity states have a charge difference $\delta Q_j$ on QD $j$. At the sweet spots in Fig.~\ref{fig_22} we find $|\delta Q_{L,R}| \approx 5 \times 10^{-3}$, with significantly larger values away from the sweet spots. In summary, based on our results above, we draw the important conclusion that it is indeed possible to find sweet spots with localized MBSs in our system.

%{\color{blue} The coupling to QD $C$ can in principle induce Yu-Shiba-Rusinov \cite{YU1965,Shiba1968,Rusinov1969} (YSR) states in QDs $L$ and $R$ but since YSR physics is compatible with poor man's MBSs, possible YSR states would not interfere with the MBSs quality. See SI \cite{SI} for a discussion on possible YSR states in QD $C$.}

{\it Transport spectroscopy.}
Next we focus on how to experimentally find the sweet spots with significant MP based on transport spectroscopy. We consider a transport setup according to Fig.~\ref{fig1}(a) with QDs $L$ and $R$ coupled with tunnel couplings $\Gamma_L = \Gamma_R = \Gamma$ to normal leads with applied voltages $V_L$ and $V_R$ (the superconductor is kept grounded). The normal leads are kept at temperature $T =  \Delta/ 40$. Focusing on the regime $\Gamma \ll T$, we calculate the current based on first-order rate equations~\cite{Kirsanskas_CPC2017, SI}. In this regime, cotunneling, Kondo correlations, and renormalization of the QD energies due to the coupling to the normal leads are negligible.

Figures~\ref{fig_3}(a), (c) and (e) show $\delta E_0$, just as in Fig.~\ref{fig_22}(a), but now as a function of $\varepsilon_L$ and $\varepsilon_R$ with $\varepsilon_C$ as in the lines marked A (a), B (c) and C (e) in Fig.~\ref{fig_22}(a). The local zero-bias conductance, $G_{jj} = dI_j/dV_j$ at $V_L = V_R = 0$, is plotted in the SI~\cite{SI} and shows peaks along the even-odd degeneracy lines. However, depending on $T$ relative to the gap to excited states, it can be hard to accurately determine the sweet spot based on a local conductance measurement. It is known that nonlocal conductance, for example $G_{LR} = dI_L/dV_R$, can reveal additional information about subgap states~\cite{Pikulin2021, Danon2020, Poeschl2022, Maiani2022}. Figures~\ref{fig_3}(b), (d) and (f) show $G_{LR}$ corresponding to the parameters in Figs.~\ref{fig_3}(a), (c) and (e). The MBS sweet spot, present only in Fig.~\ref{fig_3}(b), gives rise to a distinct $G_{LR}$ texture, with $G_{LR} = 0$ at the degeneracy lines which cross at the sweet spot, and equal magnitudes of positive and negative $G_{LR}$. In contrast, for parameters where there is no sweet spot, zeros of $G_{LR}$ do not coincide with degeneracy lines and $G_{LR}$ is dominated by either positive or negative values. This is in qualitative agreement with the experimental findings in Ref.~\cite{Dvir_arXiv2022}. 

\begin{figure}
	\includegraphics[width=0.46\textwidth,trim={0cm -0.2cm .0cm 0.cm},clip]{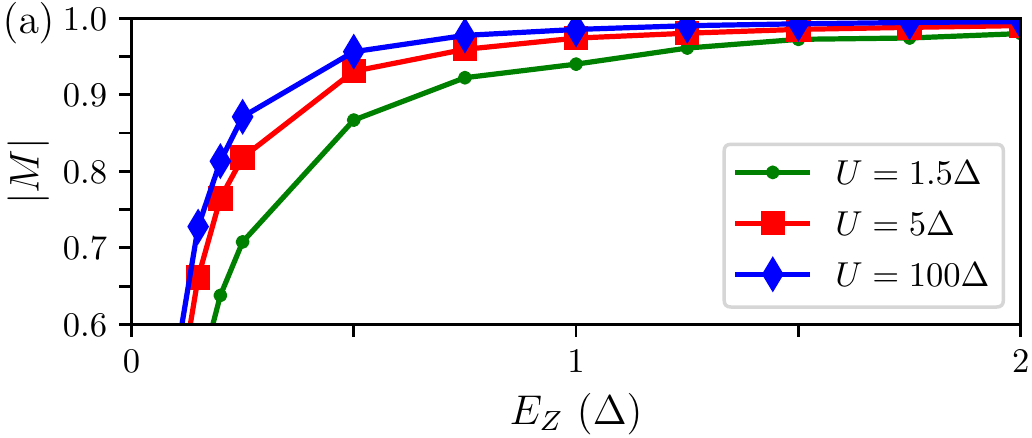}
	\includegraphics[width=0.23\textwidth,trim={0.0cm 0cm -0.0cm 0cm},clip]{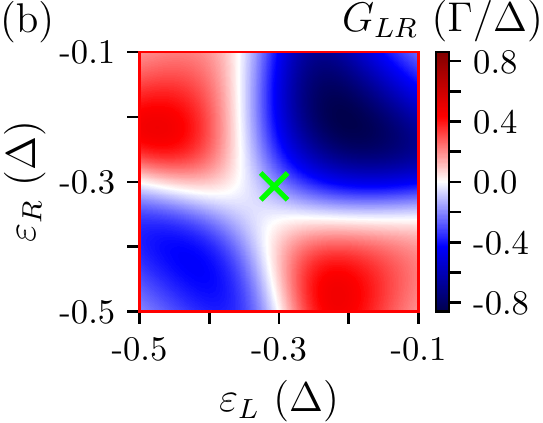}
	\includegraphics[width=0.23\textwidth,trim={0.0cm 0cm -0.0cm 0cm},clip]{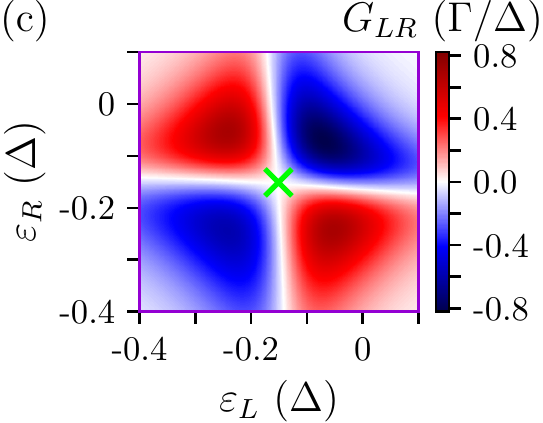}
	\caption{ (a)~$|M|$ as a function of $E_Z$ for different $U$. (b)~$G_{LR}$ as a function of $\varepsilon_L$ and $\varepsilon_R$ for $E_Z = 0.15\Delta$ and $U=5\Delta$. There is an apparent sweet spot at $\varepsilon_C \approx -0.558 \Delta$,  $\varepsilon_L = \varepsilon_R \approx -0.316 \Delta$ (marked with a green cross) with a low $|M| \approx 0.661$. (c) Same as (b) but for the same parameters as in Fig.~\ref{fig_3}(a) [zoomed in version of Fig.~\ref{fig_3}(a)]. At the sweet spot marked with the green cross in (c), $|M|\approx 0.985$.}
	\label{fig_4}
\end{figure}

{\it Low-quality MBSs.} 
Finally, we investigate how the MBS quality, as quantified by the MP, depends on the different parameters, and how we can avoid being fooled by an apparent sweet spot with low MP (``low-MP sweet spot" in the following). Figure~\ref{fig_4}(a) shows $|M|$ as a function of $E_Z = E_{ZL} = E_{ZR}$ for different values of $U$ with $\varepsilon_C$ and $\varepsilon_L = \varepsilon_R$ adjusted to an even-odd degeneracy with the highest possible $|M|$ (all other parameters are the same as above). For large $E_Z$ we find that $|M| \rightarrow 1$, which is to be expected as the model then approaches the original poor man's MBS model~\cite{Leijnse_PRB2012}. Importantly, however, we note that the values of $E_Z$ required for a good MP are much larger than the gap to the nearest excited states which is $\sim 0.15 \Delta$ in Fig.~\ref{fig_22}(c). A large $U$ helps to maintain high MP for smaller $E_Z$, which is also to be expected as it suppresses local Andreev reflection and prevents double occupation of the outer QDs.  

The appearance of low-MP sweet spots presents a challenge for experiments aiming to identify and eventually utilize MBSs. Figure~\ref{fig_4}(a) shows what to expect for a given set of parameters, but a direct experimental signature that can distinguish between high- and low-MP sweet spots would be desirable. As we show in the SI~\cite{SI}, $\delta E_0(\varepsilon_L, \varepsilon_R, \varepsilon_C)$ (and therefore the local conductances) are rather similar for high- and low-MP sweet spots. We also find a similarly small $\delta Q$ [see Eq.~(\ref{deltaQ})] for high- and low-MP sweet spots. Fortunately, a distinction can in principle be made based on a measurement of $G_{LR}$. Figure~\ref{fig_4}(b) shows $G_{LR}$ for parameters such that the MP maximum is $|M| \approx 0.661$, while Fig.~\ref{fig_4}(c) shows the same plot for parameters such that the MP maximum is $|M| \approx 0.985$. For relatively low $|M|$ the zero lines in $G_{LR}$ do not cross, and the avoided crossing does not coincide with the location of the low-MP sweet spot (marked with a green cross). The nonlocal conductance is thus finite at the degeneracy.

%%%%%%%%%%%%%%%%%%%%%%%%%%%%%%%%%%%%%%
%%%%%%%%%%%%% Conclusions %%%%%%%%%%%%%
%%%%%%%%%%%%%%%%%%%%%%%%%%%%%%%%%%%%%%

{\it Conclusions.} In this work, we have considered a system with three QDs for engineering fine-tuned MBSs, the so-called poor man's MBSs. These states require proximity-induced superconductivity on the central QD, spin-orbit coupling between the QDs, Zeeman splitting due to an external field, and fine-tuning of the energies of the QD orbitals. We have quantified the MBS quality using the MP, showing that onsite Coulomb repulsion in the outer dots and Zeeman field increase their quality. A good MBS is characterized by the simultaneous occurrence of a degenerate ground state and a high MP value for the same parameters. In contrast, a bad MBS shows low MP values at the ground state degeneracy. This characteristic leads to different nonlocal transport properties, which can be used to identify high-quality MBSs.

Although the poor man's MBSs are not topologically protected, they preserve the remaining topological properties, including the non-abelian exchange properties. For this reason, they become a promising alternative to experimentally demonstrate the exotic physics of MBSs. Proposals to measure noise \cite{Liu2015,Liu2015a,Smirnov2019,Feng2022} or entropy  \cite{Smirnov2015,Sela2019,Smirnov2021,Han2022} associated with MBSs, and to measure Majorana fusion \cite{Aasen_PRX2016,Souto_SciPost2022} and braiding \cite{Bonderson_PRL2008,Flensberg_PRL2011,van_Heck_NJP2012} are compatible with the present proposal, allowing for a definitive demonstration of the topological superconducting phase.

%%%%%%%%%%%%%%%%%%%%%%%%%%%%%%%%%%%%%%
%%%%%%%%%%%%%%% acknowledgments %%%%%%%%%%%%%%%%
%%%%%%%%%%%%%%%%%%%%%%%%%%%%%%%%%%%%%%
{\it Acknowledgments.} 
We acknowledge stimulating discussions with M. Nitsch, V. Svensson, O. A. Awoga, S. Matern, A. Danilenko, A. Pöschl, K. Flensberg, and C. M. Marcus, and funding from NanoLund, the Swedish Research Council (VR) and the European Research Council (ERC) under the European Union’s Horizon 2020 research and innovation programme under Grant Agreement No. 856526.

%\end{acknowledgments}

%\appendix

%\section{an appendix}
%Spin-orbit strength, interactions, influence of other parameters

%\section{another appendix}

%%%%%%%%%%%%%%%%%%%%%%%%%%%%%%%%%%%%%%%%%%%%%%%%%%%%%%%%%%%%%%%%%%%
%%%%%%%%%%%%%%%%%%%%%%%%%%%%%%%%%%%%%%%%%%%%%%%%%%%%%%%%%%%%%%%%%%%
%%%%%%%%%%%%%%%%%%%%%%%%%%%%%%%%%%%%%%%%%%%%%%%%%%%%%%%%%%%%%%%%%%%
%%%%%%%%%%%%%%%%%%%%%%%%%%%%%%%%%%%%%%%%%%%%%%%%%%%%%%%%%%%%%%%%%%%
%%%%%%%%%%%%%%%%%%%%%%%%%%%%%%%%%%%%%%%%%%%%%%%%%%%%%%%%%%%%%%%%%%%

%\bibliographystyle{abbrvnat}
\bibliography{poormans_coupler_so}% Produces the bibliography via BibTeX.

%apsrev4-2.bst 2019-01-14 (MD) hand-edited version of apsrev4-1.bst
%Control: key (0)
%Control: author (8) initials jnrlst
%Control: editor formatted (1) identically to author
%Control: production of article title (0) allowed
%Control: page (0) single
%Control: year (1) truncated
%Control: production of eprint (0) enabled
\begin{thebibliography}{77}%
\makeatletter
\providecommand \@ifxundefined [1]{%
 \@ifx{#1\undefined}
}%
\providecommand \@ifnum [1]{%
 \ifnum #1\expandafter \@firstoftwo
 \else \expandafter \@secondoftwo
 \fi
}%
\providecommand \@ifx [1]{%
 \ifx #1\expandafter \@firstoftwo
 \else \expandafter \@secondoftwo
 \fi
}%
\providecommand \natexlab [1]{#1}%
\providecommand \enquote  [1]{``#1''}%
\providecommand \bibnamefont  [1]{#1}%
\providecommand \bibfnamefont [1]{#1}%
\providecommand \citenamefont [1]{#1}%
\providecommand \href@noop [0]{\@secondoftwo}%
\providecommand \href [0]{\begingroup \@sanitize@url \@href}%
\providecommand \@href[1]{\@@startlink{#1}\@@href}%
\providecommand \@@href[1]{\endgroup#1\@@endlink}%
\providecommand \@sanitize@url [0]{\catcode `\\12\catcode `\$12\catcode
  `\&12\catcode `\#12\catcode `\^12\catcode `\_12\catcode `\%12\relax}%
\providecommand \@@startlink[1]{}%
\providecommand \@@endlink[0]{}%
\providecommand \url  [0]{\begingroup\@sanitize@url \@url }%
\providecommand \@url [1]{\endgroup\@href {#1}{\urlprefix }}%
\providecommand \urlprefix  [0]{URL }%
\providecommand \Eprint [0]{\href }%
\providecommand \doibase [0]{https://doi.org/}%
\providecommand \selectlanguage [0]{\@gobble}%
\providecommand \bibinfo  [0]{\@secondoftwo}%
\providecommand \bibfield  [0]{\@secondoftwo}%
\providecommand \translation [1]{[#1]}%
\providecommand \BibitemOpen [0]{}%
\providecommand \bibitemStop [0]{}%
\providecommand \bibitemNoStop [0]{.\EOS\space}%
\providecommand \EOS [0]{\spacefactor3000\relax}%
\providecommand \BibitemShut  [1]{\csname bibitem#1\endcsname}%
\let\auto@bib@innerbib\@empty
%</preamble>
\bibitem [{\citenamefont {Alicea}(2012)}]{Alicea_RPP2012}%
  \BibitemOpen
  \bibfield  {author} {\bibinfo {author} {\bibfnamefont {J.}~\bibnamefont
  {Alicea}},\ }\href@noop {} {\bibfield  {journal} {\bibinfo  {journal} {Rep.
  Prog. Phys.}\ }\textbf {\bibinfo {volume} {75}},\ \bibinfo {pages} {076501}
  (\bibinfo {year} {2012})}\BibitemShut {NoStop}%
\bibitem [{\citenamefont {Leijnse}\ and\ \citenamefont
  {Flensberg}(2012{\natexlab{a}})}]{LeijnseReview}%
  \BibitemOpen
  \bibfield  {author} {\bibinfo {author} {\bibfnamefont {M.}~\bibnamefont
  {Leijnse}}\ and\ \bibinfo {author} {\bibfnamefont {K.}~\bibnamefont
  {Flensberg}},\ }\href {https://doi.org/10.1088/0268-1242/27/12/124003}
  {\bibfield  {journal} {\bibinfo  {journal} {Semicond. Sci. Technol.}\
  }\textbf {\bibinfo {volume} {27}},\ \bibinfo {pages} {124003} (\bibinfo
  {year} {2012}{\natexlab{a}})}\BibitemShut {NoStop}%
\bibitem [{\citenamefont {Aguado}(2017)}]{AguadoReview}%
  \BibitemOpen
  \bibfield  {author} {\bibinfo {author} {\bibfnamefont {R.}~\bibnamefont
  {Aguado}},\ }\href {https://doi.org/10.1393/ncr/i2017-10141-9} {\bibfield
  {journal} {\bibinfo  {journal} {Riv. Nuovo Cimento}\ }\textbf {\bibinfo
  {volume} {40}},\ \bibinfo {pages} {523} (\bibinfo {year} {2017})}\BibitemShut
  {NoStop}%
\bibitem [{\citenamefont {Beenakker}(2020)}]{BeenakkerReview_20}%
  \BibitemOpen
  \bibfield  {author} {\bibinfo {author} {\bibfnamefont {C.~W.~J.}\
  \bibnamefont {Beenakker}},\ }\href
  {https://doi.org/10.21468/SciPostPhysLectNotes.15} {\bibfield  {journal}
  {\bibinfo  {journal} {SciPost Phys. Lect. Notes}\ ,\ \bibinfo {pages} {15}}
  (\bibinfo {year} {2020})}\BibitemShut {NoStop}%
\bibitem [{\citenamefont {Nayak}\ \emph {et~al.}(2008)\citenamefont {Nayak},
  \citenamefont {Simon}, \citenamefont {Stern}, \citenamefont {Freedman},\ and\
  \citenamefont {{Das Sarma}}}]{NayakReview}%
  \BibitemOpen
  \bibfield  {author} {\bibinfo {author} {\bibfnamefont {C.}~\bibnamefont
  {Nayak}}, \bibinfo {author} {\bibfnamefont {S.~H.}\ \bibnamefont {Simon}},
  \bibinfo {author} {\bibfnamefont {A.}~\bibnamefont {Stern}}, \bibinfo
  {author} {\bibfnamefont {M.}~\bibnamefont {Freedman}},\ and\ \bibinfo
  {author} {\bibfnamefont {S.}~\bibnamefont {{Das Sarma}}},\ }\href
  {https://doi.org/10.1103/RevModPhys.80.1083} {\bibfield  {journal} {\bibinfo
  {journal} {Rev. Mod. Phys.}\ }\textbf {\bibinfo {volume} {80}},\ \bibinfo
  {pages} {1083} (\bibinfo {year} {2008})}\BibitemShut {NoStop}%
\bibitem [{\citenamefont {Kitaev}(2001)}]{Kitaev_2001}%
  \BibitemOpen
  \bibfield  {author} {\bibinfo {author} {\bibfnamefont {A.~Y.}\ \bibnamefont
  {Kitaev}},\ }\href {https://doi.org/10.1070/1063-7869/44/10s/s29} {\bibfield
  {journal} {\bibinfo  {journal} {Phys. Usp.}\ }\textbf {\bibinfo {volume}
  {44}},\ \bibinfo {pages} {131} (\bibinfo {year} {2001})}\BibitemShut
  {NoStop}%
\bibitem [{\citenamefont {Lutchyn}\ \emph {et~al.}(2010)\citenamefont
  {Lutchyn}, \citenamefont {Sau},\ and\ \citenamefont
  {Das~Sarma}}]{Lutchyn_PRL2010}%
  \BibitemOpen
  \bibfield  {author} {\bibinfo {author} {\bibfnamefont {R.~M.}\ \bibnamefont
  {Lutchyn}}, \bibinfo {author} {\bibfnamefont {J.~D.}\ \bibnamefont {Sau}},\
  and\ \bibinfo {author} {\bibfnamefont {S.}~\bibnamefont {Das~Sarma}},\ }\href
  {https://doi.org/10.1103/PhysRevLett.105.077001} {\bibfield  {journal}
  {\bibinfo  {journal} {Phys. Rev. Lett.}\ }\textbf {\bibinfo {volume} {105}},\
  \bibinfo {pages} {077001} (\bibinfo {year} {2010})}\BibitemShut {NoStop}%
\bibitem [{\citenamefont {Oreg}\ \emph {et~al.}(2010)\citenamefont {Oreg},
  \citenamefont {Refael},\ and\ \citenamefont {von Oppen}}]{Oreg_PRL2010}%
  \BibitemOpen
  \bibfield  {author} {\bibinfo {author} {\bibfnamefont {Y.}~\bibnamefont
  {Oreg}}, \bibinfo {author} {\bibfnamefont {G.}~\bibnamefont {Refael}},\ and\
  \bibinfo {author} {\bibfnamefont {F.}~\bibnamefont {von Oppen}},\ }\href
  {https://doi.org/10.1103/PhysRevLett.105.177002} {\bibfield  {journal}
  {\bibinfo  {journal} {Phys. Rev. Lett.}\ }\textbf {\bibinfo {volume} {105}},\
  \bibinfo {pages} {177002} (\bibinfo {year} {2010})}\BibitemShut {NoStop}%
\bibitem [{\citenamefont {Nadj-Perge}\ \emph {et~al.}(2013)\citenamefont
  {Nadj-Perge}, \citenamefont {Drozdov}, \citenamefont {Bernevig},\ and\
  \citenamefont {Yazdani}}]{PhysRevB.88.020407}%
  \BibitemOpen
  \bibfield  {author} {\bibinfo {author} {\bibfnamefont {S.}~\bibnamefont
  {Nadj-Perge}}, \bibinfo {author} {\bibfnamefont {I.~K.}\ \bibnamefont
  {Drozdov}}, \bibinfo {author} {\bibfnamefont {B.~A.}\ \bibnamefont
  {Bernevig}},\ and\ \bibinfo {author} {\bibfnamefont {A.}~\bibnamefont
  {Yazdani}},\ }\href@noop {} {\bibfield  {journal} {\bibinfo  {journal} {Phys.
  Rev. B}\ }\textbf {\bibinfo {volume} {88}},\ \bibinfo {pages} {020407(R)}
  (\bibinfo {year} {2013})}\BibitemShut {NoStop}%
\bibitem [{\citenamefont {Hell}\ \emph {et~al.}(2017)\citenamefont {Hell},
  \citenamefont {Leijnse},\ and\ \citenamefont {Flensberg}}]{Hell2017}%
  \BibitemOpen
  \bibfield  {author} {\bibinfo {author} {\bibfnamefont {M.}~\bibnamefont
  {Hell}}, \bibinfo {author} {\bibfnamefont {M.}~\bibnamefont {Leijnse}},\ and\
  \bibinfo {author} {\bibfnamefont {K.}~\bibnamefont {Flensberg}},\ }\href
  {https://doi.org/10.1103/physrevlett.118.107701} {\bibfield  {journal}
  {\bibinfo  {journal} {Phys. Rev. Lett.}\ }\textbf {\bibinfo {volume} {118}},\
  \bibinfo {pages} {107701} (\bibinfo {year} {2017})}\BibitemShut {NoStop}%
\bibitem [{\citenamefont {Pientka}\ \emph {et~al.}(2017)\citenamefont
  {Pientka}, \citenamefont {Keselman}, \citenamefont {Berg}, \citenamefont
  {Yacoby}, \citenamefont {Stern},\ and\ \citenamefont
  {Halperin}}]{Pientka2017}%
  \BibitemOpen
  \bibfield  {author} {\bibinfo {author} {\bibfnamefont {F.}~\bibnamefont
  {Pientka}}, \bibinfo {author} {\bibfnamefont {A.}~\bibnamefont {Keselman}},
  \bibinfo {author} {\bibfnamefont {E.}~\bibnamefont {Berg}}, \bibinfo {author}
  {\bibfnamefont {A.}~\bibnamefont {Yacoby}}, \bibinfo {author} {\bibfnamefont
  {A.}~\bibnamefont {Stern}},\ and\ \bibinfo {author} {\bibfnamefont {B.~I.}\
  \bibnamefont {Halperin}},\ }\href@noop {} {\bibfield  {journal} {\bibinfo
  {journal} {Phys. Rev. X}\ }\textbf {\bibinfo {volume} {7}},\ \bibinfo {pages}
  {021032} (\bibinfo {year} {2017})}\BibitemShut {NoStop}%
\bibitem [{\citenamefont {Vaitiek{\.{e}}nas}\ \emph {et~al.}(2020)\citenamefont
  {Vaitiek{\.{e}}nas}, \citenamefont {Winkler}, \citenamefont {van Heck},
  \citenamefont {Karzig}, \citenamefont {Deng}, \citenamefont {Flensberg},
  \citenamefont {Glazman}, \citenamefont {Nayak}, \citenamefont {Krogstrup},
  \citenamefont {Lutchyn},\ and\ \citenamefont {Marcus}}]{Vaitiekenas2020}%
  \BibitemOpen
  \bibfield  {author} {\bibinfo {author} {\bibfnamefont {S.}~\bibnamefont
  {Vaitiek{\.{e}}nas}}, \bibinfo {author} {\bibfnamefont {G.~W.}\ \bibnamefont
  {Winkler}}, \bibinfo {author} {\bibfnamefont {B.}~\bibnamefont {van Heck}},
  \bibinfo {author} {\bibfnamefont {T.}~\bibnamefont {Karzig}}, \bibinfo
  {author} {\bibfnamefont {M.-T.}\ \bibnamefont {Deng}}, \bibinfo {author}
  {\bibfnamefont {K.}~\bibnamefont {Flensberg}}, \bibinfo {author}
  {\bibfnamefont {L.~I.}\ \bibnamefont {Glazman}}, \bibinfo {author}
  {\bibfnamefont {C.}~\bibnamefont {Nayak}}, \bibinfo {author} {\bibfnamefont
  {P.}~\bibnamefont {Krogstrup}}, \bibinfo {author} {\bibfnamefont {R.~M.}\
  \bibnamefont {Lutchyn}},\ and\ \bibinfo {author} {\bibfnamefont {C.~M.}\
  \bibnamefont {Marcus}},\ }\href@noop {} {\bibfield  {journal} {\bibinfo
  {journal} {Science}\ }\textbf {\bibinfo {volume} {367}} (\bibinfo {year}
  {2020})}\BibitemShut {NoStop}%
\bibitem [{\citenamefont {Flensberg}\ \emph {et~al.}(2021)\citenamefont
  {Flensberg}, \citenamefont {von Oppen},\ and\ \citenamefont
  {Stern}}]{flensberg2021engineered}%
  \BibitemOpen
  \bibfield  {author} {\bibinfo {author} {\bibfnamefont {K.}~\bibnamefont
  {Flensberg}}, \bibinfo {author} {\bibfnamefont {F.}~\bibnamefont {von
  Oppen}},\ and\ \bibinfo {author} {\bibfnamefont {A.}~\bibnamefont {Stern}},\
  }\href {https://doi.org/https://doi.org/10.1038/s41578-021-00336-6}
  {\bibfield  {journal} {\bibinfo  {journal} {Nat. Rev. Mater.}\ }\textbf
  {\bibinfo {volume} {6}},\ \bibinfo {pages} {944} (\bibinfo {year}
  {2021})}\BibitemShut {NoStop}%
\bibitem [{\citenamefont {Mourik}\ \emph {et~al.}(2012)\citenamefont {Mourik},
  \citenamefont {Zuo}, \citenamefont {Frolov}, \citenamefont {Plissard},
  \citenamefont {Bakkers},\ and\ \citenamefont
  {Kouwenhoven}}]{Mourik_science2012}%
  \BibitemOpen
  \bibfield  {author} {\bibinfo {author} {\bibfnamefont {V.}~\bibnamefont
  {Mourik}}, \bibinfo {author} {\bibfnamefont {K.}~\bibnamefont {Zuo}},
  \bibinfo {author} {\bibfnamefont {S.~M.}\ \bibnamefont {Frolov}}, \bibinfo
  {author} {\bibfnamefont {S.~R.}\ \bibnamefont {Plissard}}, \bibinfo {author}
  {\bibfnamefont {E.~P. A.~M.}\ \bibnamefont {Bakkers}},\ and\ \bibinfo
  {author} {\bibfnamefont {L.~P.}\ \bibnamefont {Kouwenhoven}},\ }\href
  {https://doi.org/10.1126/science.1222360} {\bibfield  {journal} {\bibinfo
  {journal} {Science}\ }\textbf {\bibinfo {volume} {336}},\ \bibinfo {pages}
  {1003} (\bibinfo {year} {2012})}\BibitemShut {NoStop}%
\bibitem [{\citenamefont {Deng}\ \emph {et~al.}(2012)\citenamefont {Deng},
  \citenamefont {Yu}, \citenamefont {Huang}, \citenamefont {Larsson},
  \citenamefont {Caroff},\ and\ \citenamefont {Xu}}]{deng2012anomalous}%
  \BibitemOpen
  \bibfield  {author} {\bibinfo {author} {\bibfnamefont {M.~T.}\ \bibnamefont
  {Deng}}, \bibinfo {author} {\bibfnamefont {C.~L.}\ \bibnamefont {Yu}},
  \bibinfo {author} {\bibfnamefont {G.~Y.}\ \bibnamefont {Huang}}, \bibinfo
  {author} {\bibfnamefont {M.}~\bibnamefont {Larsson}}, \bibinfo {author}
  {\bibfnamefont {P.}~\bibnamefont {Caroff}},\ and\ \bibinfo {author}
  {\bibfnamefont {H.~Q.}\ \bibnamefont {Xu}},\ }\href@noop {} {\bibfield
  {journal} {\bibinfo  {journal} {Nano Lett.}\ }\textbf {\bibinfo {volume}
  {12}},\ \bibinfo {pages} {6414} (\bibinfo {year} {2012})}\BibitemShut
  {NoStop}%
\bibitem [{\citenamefont {Finck}\ \emph {et~al.}(2013)\citenamefont {Finck},
  \citenamefont {Van~Harlingen}, \citenamefont {Mohseni}, \citenamefont
  {Jung},\ and\ \citenamefont {Li}}]{finck2013anomalous}%
  \BibitemOpen
  \bibfield  {author} {\bibinfo {author} {\bibfnamefont {A.~D.~K.}\
  \bibnamefont {Finck}}, \bibinfo {author} {\bibfnamefont {D.~J.}\ \bibnamefont
  {Van~Harlingen}}, \bibinfo {author} {\bibfnamefont {P.~K.}\ \bibnamefont
  {Mohseni}}, \bibinfo {author} {\bibfnamefont {K.}~\bibnamefont {Jung}},\ and\
  \bibinfo {author} {\bibfnamefont {X.}~\bibnamefont {Li}},\ }\href
  {https://doi.org/10.1103/PhysRevLett.110.126406} {\bibfield  {journal}
  {\bibinfo  {journal} {Phys. Rev. Lett.}\ }\textbf {\bibinfo {volume} {110}},\
  \bibinfo {pages} {126406} (\bibinfo {year} {2013})}\BibitemShut {NoStop}%
\bibitem [{\citenamefont {Nadj-Perge}\ \emph {et~al.}(2014)\citenamefont
  {Nadj-Perge}, \citenamefont {Drozdov}, \citenamefont {Li}, \citenamefont
  {Chen}, \citenamefont {Jeon}, \citenamefont {Seo}, \citenamefont {MacDonald},
  \citenamefont {Bernevig},\ and\ \citenamefont {Yazdani}}]{NadjPerge2014}%
  \BibitemOpen
  \bibfield  {author} {\bibinfo {author} {\bibfnamefont {S.}~\bibnamefont
  {Nadj-Perge}}, \bibinfo {author} {\bibfnamefont {I.~K.}\ \bibnamefont
  {Drozdov}}, \bibinfo {author} {\bibfnamefont {J.}~\bibnamefont {Li}},
  \bibinfo {author} {\bibfnamefont {H.}~\bibnamefont {Chen}}, \bibinfo {author}
  {\bibfnamefont {S.}~\bibnamefont {Jeon}}, \bibinfo {author} {\bibfnamefont
  {J.}~\bibnamefont {Seo}}, \bibinfo {author} {\bibfnamefont {A.~H.}\
  \bibnamefont {MacDonald}}, \bibinfo {author} {\bibfnamefont {B.~A.}\
  \bibnamefont {Bernevig}},\ and\ \bibinfo {author} {\bibfnamefont
  {A.}~\bibnamefont {Yazdani}},\ }\href@noop {} {\bibfield  {journal} {\bibinfo
   {journal} {Science}\ }\textbf {\bibinfo {volume} {346}},\ \bibinfo {pages}
  {602} (\bibinfo {year} {2014})}\BibitemShut {NoStop}%
\bibitem [{\citenamefont {Deng}\ \emph {et~al.}(2016)\citenamefont {Deng},
  \citenamefont {Vaitiekėnas}, \citenamefont {Hansen}, \citenamefont {Danon},
  \citenamefont {Leijnse}, \citenamefont {Flensberg}, \citenamefont {Nygård},
  \citenamefont {Krogstrup},\ and\ \citenamefont {Marcus}}]{deng2016majorana}%
  \BibitemOpen
  \bibfield  {author} {\bibinfo {author} {\bibfnamefont {M.~T.}\ \bibnamefont
  {Deng}}, \bibinfo {author} {\bibfnamefont {S.}~\bibnamefont {Vaitiekėnas}},
  \bibinfo {author} {\bibfnamefont {E.~B.}\ \bibnamefont {Hansen}}, \bibinfo
  {author} {\bibfnamefont {J.}~\bibnamefont {Danon}}, \bibinfo {author}
  {\bibfnamefont {M.}~\bibnamefont {Leijnse}}, \bibinfo {author} {\bibfnamefont
  {K.}~\bibnamefont {Flensberg}}, \bibinfo {author} {\bibfnamefont
  {J.}~\bibnamefont {Nygård}}, \bibinfo {author} {\bibfnamefont
  {P.}~\bibnamefont {Krogstrup}},\ and\ \bibinfo {author} {\bibfnamefont
  {C.~M.}\ \bibnamefont {Marcus}},\ }\href@noop {} {\bibfield  {journal}
  {\bibinfo  {journal} {Science}\ }\textbf {\bibinfo {volume} {354}},\ \bibinfo
  {pages} {1557} (\bibinfo {year} {2016})}\BibitemShut {NoStop}%
\bibitem [{\citenamefont {Nichele}\ \emph {et~al.}(2017)\citenamefont
  {Nichele}, \citenamefont {Drachmann}, \citenamefont {Whiticar}, \citenamefont
  {O'Farrell}, \citenamefont {Suominen}, \citenamefont {Fornieri},
  \citenamefont {Wang}, \citenamefont {Gardner}, \citenamefont {Thomas},
  \citenamefont {Hatke}, \citenamefont {Krogstrup}, \citenamefont {Manfra},
  \citenamefont {Flensberg},\ and\ \citenamefont {Marcus}}]{Nichele_PRL2017}%
  \BibitemOpen
  \bibfield  {author} {\bibinfo {author} {\bibfnamefont {F.}~\bibnamefont
  {Nichele}}, \bibinfo {author} {\bibfnamefont {A.~C.~C.}\ \bibnamefont
  {Drachmann}}, \bibinfo {author} {\bibfnamefont {A.~M.}\ \bibnamefont
  {Whiticar}}, \bibinfo {author} {\bibfnamefont {E.~C.~T.}\ \bibnamefont
  {O'Farrell}}, \bibinfo {author} {\bibfnamefont {H.~J.}\ \bibnamefont
  {Suominen}}, \bibinfo {author} {\bibfnamefont {A.}~\bibnamefont {Fornieri}},
  \bibinfo {author} {\bibfnamefont {T.}~\bibnamefont {Wang}}, \bibinfo {author}
  {\bibfnamefont {G.~C.}\ \bibnamefont {Gardner}}, \bibinfo {author}
  {\bibfnamefont {C.}~\bibnamefont {Thomas}}, \bibinfo {author} {\bibfnamefont
  {A.~T.}\ \bibnamefont {Hatke}}, \bibinfo {author} {\bibfnamefont
  {P.}~\bibnamefont {Krogstrup}}, \bibinfo {author} {\bibfnamefont {M.~J.}\
  \bibnamefont {Manfra}}, \bibinfo {author} {\bibfnamefont {K.}~\bibnamefont
  {Flensberg}},\ and\ \bibinfo {author} {\bibfnamefont {C.~M.}\ \bibnamefont
  {Marcus}},\ }\href {https://doi.org/10.1103/PhysRevLett.119.136803}
  {\bibfield  {journal} {\bibinfo  {journal} {Phys. Rev. Lett.}\ }\textbf
  {\bibinfo {volume} {119}},\ \bibinfo {pages} {136803} (\bibinfo {year}
  {2017})}\BibitemShut {NoStop}%
\bibitem [{\citenamefont {Lutchyn}\ \emph {et~al.}(2018)\citenamefont
  {Lutchyn}, \citenamefont {Bakkers}, \citenamefont {Kouwenhoven},
  \citenamefont {Krogstrup}, \citenamefont {Marcus},\ and\ \citenamefont
  {Oreg}}]{lutchyn2018majorana}%
  \BibitemOpen
  \bibfield  {author} {\bibinfo {author} {\bibfnamefont {R.~M.}\ \bibnamefont
  {Lutchyn}}, \bibinfo {author} {\bibfnamefont {E.~P.}\ \bibnamefont
  {Bakkers}}, \bibinfo {author} {\bibfnamefont {L.~P.}\ \bibnamefont
  {Kouwenhoven}}, \bibinfo {author} {\bibfnamefont {P.}~\bibnamefont
  {Krogstrup}}, \bibinfo {author} {\bibfnamefont {C.~M.}\ \bibnamefont
  {Marcus}},\ and\ \bibinfo {author} {\bibfnamefont {Y.}~\bibnamefont {Oreg}},\
  }\href {https://doi.org/https://doi.org/10.1038/s41578-018-0003-1} {\bibfield
   {journal} {\bibinfo  {journal} {Nat. Rev. Mater.}\ }\textbf {\bibinfo
  {volume} {3}},\ \bibinfo {pages} {52} (\bibinfo {year} {2018})}\BibitemShut
  {NoStop}%
\bibitem [{\citenamefont {Fornieri}\ \emph {et~al.}(2019)\citenamefont
  {Fornieri}, \citenamefont {Whiticar}, \citenamefont {Setiawan}, \citenamefont
  {Portol{\'{e}}s}, \citenamefont {Drachmann}, \citenamefont {Keselman},
  \citenamefont {Gronin}, \citenamefont {Thomas}, \citenamefont {Wang},
  \citenamefont {Kallaher}, \citenamefont {Gardner}, \citenamefont {Berg},
  \citenamefont {Manfra}, \citenamefont {Stern}, \citenamefont {Marcus},\ and\
  \citenamefont {Nichele}}]{Fornieri2019}%
  \BibitemOpen
  \bibfield  {author} {\bibinfo {author} {\bibfnamefont {A.}~\bibnamefont
  {Fornieri}}, \bibinfo {author} {\bibfnamefont {A.~M.}\ \bibnamefont
  {Whiticar}}, \bibinfo {author} {\bibfnamefont {F.}~\bibnamefont {Setiawan}},
  \bibinfo {author} {\bibfnamefont {E.}~\bibnamefont {Portol{\'{e}}s}},
  \bibinfo {author} {\bibfnamefont {A.~C.~C.}\ \bibnamefont {Drachmann}},
  \bibinfo {author} {\bibfnamefont {A.}~\bibnamefont {Keselman}}, \bibinfo
  {author} {\bibfnamefont {S.}~\bibnamefont {Gronin}}, \bibinfo {author}
  {\bibfnamefont {C.}~\bibnamefont {Thomas}}, \bibinfo {author} {\bibfnamefont
  {T.}~\bibnamefont {Wang}}, \bibinfo {author} {\bibfnamefont {R.}~\bibnamefont
  {Kallaher}}, \bibinfo {author} {\bibfnamefont {G.~C.}\ \bibnamefont
  {Gardner}}, \bibinfo {author} {\bibfnamefont {E.}~\bibnamefont {Berg}},
  \bibinfo {author} {\bibfnamefont {M.~J.}\ \bibnamefont {Manfra}}, \bibinfo
  {author} {\bibfnamefont {A.}~\bibnamefont {Stern}}, \bibinfo {author}
  {\bibfnamefont {C.~M.}\ \bibnamefont {Marcus}},\ and\ \bibinfo {author}
  {\bibfnamefont {F.}~\bibnamefont {Nichele}},\ }\href@noop {} {\bibfield
  {journal} {\bibinfo  {journal} {Nature}\ }\textbf {\bibinfo {volume} {569}},\
  \bibinfo {pages} {89} (\bibinfo {year} {2019})}\BibitemShut {NoStop}%
\bibitem [{\citenamefont {Ren}\ \emph {et~al.}(2019)\citenamefont {Ren},
  \citenamefont {Pientka}, \citenamefont {Hart}, \citenamefont {Pierce},
  \citenamefont {Kosowsky}, \citenamefont {Lunczer}, \citenamefont {Schlereth},
  \citenamefont {Scharf}, \citenamefont {Hankiewicz}, \citenamefont
  {Molenkamp}, \citenamefont {Halperin},\ and\ \citenamefont
  {Yacoby}}]{Ren2019}%
  \BibitemOpen
  \bibfield  {author} {\bibinfo {author} {\bibfnamefont {H.}~\bibnamefont
  {Ren}}, \bibinfo {author} {\bibfnamefont {F.}~\bibnamefont {Pientka}},
  \bibinfo {author} {\bibfnamefont {S.}~\bibnamefont {Hart}}, \bibinfo {author}
  {\bibfnamefont {A.~T.}\ \bibnamefont {Pierce}}, \bibinfo {author}
  {\bibfnamefont {M.}~\bibnamefont {Kosowsky}}, \bibinfo {author}
  {\bibfnamefont {L.}~\bibnamefont {Lunczer}}, \bibinfo {author} {\bibfnamefont
  {R.}~\bibnamefont {Schlereth}}, \bibinfo {author} {\bibfnamefont
  {B.}~\bibnamefont {Scharf}}, \bibinfo {author} {\bibfnamefont {E.~M.}\
  \bibnamefont {Hankiewicz}}, \bibinfo {author} {\bibfnamefont {L.~W.}\
  \bibnamefont {Molenkamp}}, \bibinfo {author} {\bibfnamefont {B.~I.}\
  \bibnamefont {Halperin}},\ and\ \bibinfo {author} {\bibfnamefont
  {A.}~\bibnamefont {Yacoby}},\ }\href@noop {} {\bibfield  {journal} {\bibinfo
  {journal} {Nature}\ }\textbf {\bibinfo {volume} {569}},\ \bibinfo {pages}
  {93} (\bibinfo {year} {2019})}\BibitemShut {NoStop}%
\bibitem [{\citenamefont {Prada}\ \emph {et~al.}(2012)\citenamefont {Prada},
  \citenamefont {San-Jose},\ and\ \citenamefont {Aguado}}]{Prada_PRB2012}%
  \BibitemOpen
  \bibfield  {author} {\bibinfo {author} {\bibfnamefont {E.}~\bibnamefont
  {Prada}}, \bibinfo {author} {\bibfnamefont {P.}~\bibnamefont {San-Jose}},\
  and\ \bibinfo {author} {\bibfnamefont {R.}~\bibnamefont {Aguado}},\ }\href
  {https://doi.org/10.1103/PhysRevB.86.180503} {\bibfield  {journal} {\bibinfo
  {journal} {Phys. Rev. B}\ }\textbf {\bibinfo {volume} {86}},\ \bibinfo
  {pages} {180503(R)} (\bibinfo {year} {2012})}\BibitemShut {NoStop}%
\bibitem [{\citenamefont {Kells}\ \emph {et~al.}(2012)\citenamefont {Kells},
  \citenamefont {Meidan},\ and\ \citenamefont {Brouwer}}]{Kells_PRB12}%
  \BibitemOpen
  \bibfield  {author} {\bibinfo {author} {\bibfnamefont {G.}~\bibnamefont
  {Kells}}, \bibinfo {author} {\bibfnamefont {D.}~\bibnamefont {Meidan}},\ and\
  \bibinfo {author} {\bibfnamefont {P.~W.}\ \bibnamefont {Brouwer}},\ }\href
  {https://doi.org/10.1103/PhysRevB.86.100503} {\bibfield  {journal} {\bibinfo
  {journal} {Phys. Rev. B}\ }\textbf {\bibinfo {volume} {86}},\ \bibinfo
  {pages} {100503(R)} (\bibinfo {year} {2012})}\BibitemShut {NoStop}%
\bibitem [{\citenamefont {Liu}\ \emph {et~al.}(2012)\citenamefont {Liu},
  \citenamefont {Potter}, \citenamefont {Law},\ and\ \citenamefont
  {Lee}}]{Liu2012}%
  \BibitemOpen
  \bibfield  {author} {\bibinfo {author} {\bibfnamefont {J.}~\bibnamefont
  {Liu}}, \bibinfo {author} {\bibfnamefont {A.~C.}\ \bibnamefont {Potter}},
  \bibinfo {author} {\bibfnamefont {K.~T.}\ \bibnamefont {Law}},\ and\ \bibinfo
  {author} {\bibfnamefont {P.~A.}\ \bibnamefont {Lee}},\ }\href
  {https://doi.org/10.1103/physrevlett.109.267002} {\bibfield  {journal}
  {\bibinfo  {journal} {Phys. Rev. Lett.}\ }\textbf {\bibinfo {volume} {109}},\
  \bibinfo {pages} {267002} (\bibinfo {year} {2012})}\BibitemShut {NoStop}%
\bibitem [{\citenamefont {Liu}\ \emph {et~al.}(2017)\citenamefont {Liu},
  \citenamefont {Sau}, \citenamefont {Stanescu},\ and\ \citenamefont
  {Das~Sarma}}]{Liu2017}%
  \BibitemOpen
  \bibfield  {author} {\bibinfo {author} {\bibfnamefont {C.-X.}\ \bibnamefont
  {Liu}}, \bibinfo {author} {\bibfnamefont {J.~D.}\ \bibnamefont {Sau}},
  \bibinfo {author} {\bibfnamefont {T.~D.}\ \bibnamefont {Stanescu}},\ and\
  \bibinfo {author} {\bibfnamefont {S.}~\bibnamefont {Das~Sarma}},\ }\href
  {https://doi.org/10.1103/physrevb.96.075161} {\bibfield  {journal} {\bibinfo
  {journal} {Phys. Rev. B}\ }\textbf {\bibinfo {volume} {96}},\ \bibinfo
  {pages} {075161} (\bibinfo {year} {2017})}\BibitemShut {NoStop}%
\bibitem [{\citenamefont {Moore}\ \emph {et~al.}(2018)\citenamefont {Moore},
  \citenamefont {Zeng}, \citenamefont {Stanescu},\ and\ \citenamefont
  {Tewari}}]{Moore_PRB18}%
  \BibitemOpen
  \bibfield  {author} {\bibinfo {author} {\bibfnamefont {C.}~\bibnamefont
  {Moore}}, \bibinfo {author} {\bibfnamefont {C.}~\bibnamefont {Zeng}},
  \bibinfo {author} {\bibfnamefont {T.~D.}\ \bibnamefont {Stanescu}},\ and\
  \bibinfo {author} {\bibfnamefont {S.}~\bibnamefont {Tewari}},\ }\href
  {https://doi.org/10.1103/PhysRevB.98.155314} {\bibfield  {journal} {\bibinfo
  {journal} {Phys. Rev. B}\ }\textbf {\bibinfo {volume} {98}},\ \bibinfo
  {pages} {155314} (\bibinfo {year} {2018})}\BibitemShut {NoStop}%
\bibitem [{\citenamefont {Reeg}\ \emph {et~al.}(2018)\citenamefont {Reeg},
  \citenamefont {Dmytruk}, \citenamefont {Chevallier}, \citenamefont {Loss},\
  and\ \citenamefont {Klinovaja}}]{reeg2018zero}%
  \BibitemOpen
  \bibfield  {author} {\bibinfo {author} {\bibfnamefont {C.}~\bibnamefont
  {Reeg}}, \bibinfo {author} {\bibfnamefont {O.}~\bibnamefont {Dmytruk}},
  \bibinfo {author} {\bibfnamefont {D.}~\bibnamefont {Chevallier}}, \bibinfo
  {author} {\bibfnamefont {D.}~\bibnamefont {Loss}},\ and\ \bibinfo {author}
  {\bibfnamefont {J.}~\bibnamefont {Klinovaja}},\ }\href
  {https://doi.org/10.1103/PhysRevB.98.245407} {\bibfield  {journal} {\bibinfo
  {journal} {Phys. Rev. B}\ }\textbf {\bibinfo {volume} {98}},\ \bibinfo
  {pages} {245407} (\bibinfo {year} {2018})}\BibitemShut {NoStop}%
\bibitem [{\citenamefont {Awoga}\ \emph {et~al.}(2019)\citenamefont {Awoga},
  \citenamefont {Cayao},\ and\ \citenamefont {Black-Schaffer}}]{Awoga_PRL2019}%
  \BibitemOpen
  \bibfield  {author} {\bibinfo {author} {\bibfnamefont {O.~A.}\ \bibnamefont
  {Awoga}}, \bibinfo {author} {\bibfnamefont {J.}~\bibnamefont {Cayao}},\ and\
  \bibinfo {author} {\bibfnamefont {A.~M.}\ \bibnamefont {Black-Schaffer}},\
  }\href {https://doi.org/10.1103/PhysRevLett.123.117001} {\bibfield  {journal}
  {\bibinfo  {journal} {Phys. Rev. Lett.}\ }\textbf {\bibinfo {volume} {123}},\
  \bibinfo {pages} {117001} (\bibinfo {year} {2019})}\BibitemShut {NoStop}%
\bibitem [{\citenamefont {Vuik}\ \emph {et~al.}(2019)\citenamefont {Vuik},
  \citenamefont {Nijholt}, \citenamefont {Akhmerov},\ and\ \citenamefont
  {Wimmer}}]{Vuik_SciPost19}%
  \BibitemOpen
  \bibfield  {author} {\bibinfo {author} {\bibfnamefont {A.}~\bibnamefont
  {Vuik}}, \bibinfo {author} {\bibfnamefont {B.}~\bibnamefont {Nijholt}},
  \bibinfo {author} {\bibfnamefont {A.~R.}\ \bibnamefont {Akhmerov}},\ and\
  \bibinfo {author} {\bibfnamefont {M.}~\bibnamefont {Wimmer}},\ }\href
  {https://doi.org/10.21468/SciPostPhys.7.5.061} {\bibfield  {journal}
  {\bibinfo  {journal} {SciPost Phys.}\ }\textbf {\bibinfo {volume} {7}},\
  \bibinfo {pages} {61} (\bibinfo {year} {2019})}\BibitemShut {NoStop}%
\bibitem [{\citenamefont {Pan}\ and\ \citenamefont
  {Das~Sarma}(2020)}]{Pan_PRR20}%
  \BibitemOpen
  \bibfield  {author} {\bibinfo {author} {\bibfnamefont {H.}~\bibnamefont
  {Pan}}\ and\ \bibinfo {author} {\bibfnamefont {S.}~\bibnamefont
  {Das~Sarma}},\ }\href {https://doi.org/10.1103/PhysRevResearch.2.013377}
  {\bibfield  {journal} {\bibinfo  {journal} {Phys. Rev. Res.}\ }\textbf
  {\bibinfo {volume} {2}},\ \bibinfo {pages} {013377} (\bibinfo {year}
  {2020})}\BibitemShut {NoStop}%
\bibitem [{\citenamefont {Prada}\ \emph {et~al.}(2020)\citenamefont {Prada},
  \citenamefont {San-Jose}, \citenamefont {de~Moor}, \citenamefont {Geresdi},
  \citenamefont {Lee}, \citenamefont {Klinovaja}, \citenamefont {Loss},
  \citenamefont {Nyg{\aa}rd}, \citenamefont {Aguado},\ and\ \citenamefont
  {Kouwenhoven}}]{Prada_review}%
  \BibitemOpen
  \bibfield  {author} {\bibinfo {author} {\bibfnamefont {E.}~\bibnamefont
  {Prada}}, \bibinfo {author} {\bibfnamefont {P.}~\bibnamefont {San-Jose}},
  \bibinfo {author} {\bibfnamefont {M.~W.~A.}\ \bibnamefont {de~Moor}},
  \bibinfo {author} {\bibfnamefont {A.}~\bibnamefont {Geresdi}}, \bibinfo
  {author} {\bibfnamefont {E.~J.~H.}\ \bibnamefont {Lee}}, \bibinfo {author}
  {\bibfnamefont {J.}~\bibnamefont {Klinovaja}}, \bibinfo {author}
  {\bibfnamefont {D.}~\bibnamefont {Loss}}, \bibinfo {author} {\bibfnamefont
  {J.}~\bibnamefont {Nyg{\aa}rd}}, \bibinfo {author} {\bibfnamefont
  {R.}~\bibnamefont {Aguado}},\ and\ \bibinfo {author} {\bibfnamefont {L.~P.}\
  \bibnamefont {Kouwenhoven}},\ }\href@noop {} {\bibfield  {journal} {\bibinfo
  {journal} {Nature Rev. Phys.}\ }\textbf {\bibinfo {volume} {2}},\ \bibinfo
  {pages} {575} (\bibinfo {year} {2020})}\BibitemShut {NoStop}%
\bibitem [{\citenamefont {Hess}\ \emph {et~al.}(2021)\citenamefont {Hess},
  \citenamefont {Legg}, \citenamefont {Loss},\ and\ \citenamefont
  {Klinovaja}}]{hess2021local}%
  \BibitemOpen
  \bibfield  {author} {\bibinfo {author} {\bibfnamefont {R.}~\bibnamefont
  {Hess}}, \bibinfo {author} {\bibfnamefont {H.~F.}\ \bibnamefont {Legg}},
  \bibinfo {author} {\bibfnamefont {D.}~\bibnamefont {Loss}},\ and\ \bibinfo
  {author} {\bibfnamefont {J.}~\bibnamefont {Klinovaja}},\ }\href
  {https://doi.org/10.1103/PhysRevB.104.075405} {\bibfield  {journal} {\bibinfo
   {journal} {Phys. Rev. B}\ }\textbf {\bibinfo {volume} {104}},\ \bibinfo
  {pages} {075405} (\bibinfo {year} {2021})}\BibitemShut {NoStop}%
\bibitem [{\citenamefont {Sau}\ and\ \citenamefont
  {Sarma}(2012)}]{Sau_NatComm2012}%
  \BibitemOpen
  \bibfield  {author} {\bibinfo {author} {\bibfnamefont {J.~D.}\ \bibnamefont
  {Sau}}\ and\ \bibinfo {author} {\bibfnamefont {S.~D.}\ \bibnamefont
  {Sarma}},\ }\href {https://doi.org/10.1038/ncomms1966} {\bibfield  {journal}
  {\bibinfo  {journal} {Nature Commun.}\ }\textbf {\bibinfo {volume} {3}},\
  \bibinfo {pages} {964} (\bibinfo {year} {2012})}\BibitemShut {NoStop}%
\bibitem [{\citenamefont {Leijnse}\ and\ \citenamefont
  {Flensberg}(2012{\natexlab{b}})}]{Leijnse_PRB2012}%
  \BibitemOpen
  \bibfield  {author} {\bibinfo {author} {\bibfnamefont {M.}~\bibnamefont
  {Leijnse}}\ and\ \bibinfo {author} {\bibfnamefont {K.}~\bibnamefont
  {Flensberg}},\ }\href {https://doi.org/10.1103/PhysRevB.86.134528} {\bibfield
   {journal} {\bibinfo  {journal} {Phys. Rev. B}\ }\textbf {\bibinfo {volume}
  {86}},\ \bibinfo {pages} {134528} (\bibinfo {year}
  {2012}{\natexlab{b}})}\BibitemShut {NoStop}%
\bibitem [{\citenamefont {Recher}\ \emph {et~al.}(2001)\citenamefont {Recher},
  \citenamefont {Sukhorukov},\ and\ \citenamefont {Loss}}]{Recher_PRB2001}%
  \BibitemOpen
  \bibfield  {author} {\bibinfo {author} {\bibfnamefont {P.}~\bibnamefont
  {Recher}}, \bibinfo {author} {\bibfnamefont {E.~V.}\ \bibnamefont
  {Sukhorukov}},\ and\ \bibinfo {author} {\bibfnamefont {D.}~\bibnamefont
  {Loss}},\ }\href {https://doi.org/10.1103/PhysRevB.63.165314} {\bibfield
  {journal} {\bibinfo  {journal} {Phys. Rev. B}\ }\textbf {\bibinfo {volume}
  {63}},\ \bibinfo {pages} {165314} (\bibinfo {year} {2001})}\BibitemShut
  {NoStop}%
\bibitem [{\citenamefont {Hofstetter}\ \emph {et~al.}(2009)\citenamefont
  {Hofstetter}, \citenamefont {Csonka}, \citenamefont {Nyg{\aa}rd},\ and\
  \citenamefont {Sch{\"o}nenberger}}]{Hofstetter_Nature2009}%
  \BibitemOpen
  \bibfield  {author} {\bibinfo {author} {\bibfnamefont {L.}~\bibnamefont
  {Hofstetter}}, \bibinfo {author} {\bibfnamefont {S.}~\bibnamefont {Csonka}},
  \bibinfo {author} {\bibfnamefont {J.}~\bibnamefont {Nyg{\aa}rd}},\ and\
  \bibinfo {author} {\bibfnamefont {C.}~\bibnamefont {Sch{\"o}nenberger}},\
  }\href {https://doi.org/10.1038/nature08432} {\bibfield  {journal} {\bibinfo
  {journal} {Nature}\ }\textbf {\bibinfo {volume} {461}},\ \bibinfo {pages}
  {960} (\bibinfo {year} {2009})}\BibitemShut {NoStop}%
\bibitem [{\citenamefont {Herrmann}\ \emph {et~al.}(2010)\citenamefont
  {Herrmann}, \citenamefont {Portier}, \citenamefont {Roche}, \citenamefont
  {Yeyati}, \citenamefont {Kontos},\ and\ \citenamefont
  {Strunk}}]{Herrmann_PRL2010}%
  \BibitemOpen
  \bibfield  {author} {\bibinfo {author} {\bibfnamefont {L.~G.}\ \bibnamefont
  {Herrmann}}, \bibinfo {author} {\bibfnamefont {F.}~\bibnamefont {Portier}},
  \bibinfo {author} {\bibfnamefont {P.}~\bibnamefont {Roche}}, \bibinfo
  {author} {\bibfnamefont {A.~L.}\ \bibnamefont {Yeyati}}, \bibinfo {author}
  {\bibfnamefont {T.}~\bibnamefont {Kontos}},\ and\ \bibinfo {author}
  {\bibfnamefont {C.}~\bibnamefont {Strunk}},\ }\href
  {https://doi.org/10.1103/PhysRevLett.104.026801} {\bibfield  {journal}
  {\bibinfo  {journal} {Phys. Rev. Lett.}\ }\textbf {\bibinfo {volume} {104}},\
  \bibinfo {pages} {026801} (\bibinfo {year} {2010})}\BibitemShut {NoStop}%
\bibitem [{\citenamefont {F\"ul\"op}\ \emph {et~al.}(2015)\citenamefont
  {F\"ul\"op}, \citenamefont {Dom\'{\i}nguez}, \citenamefont {d'Hollosy},
  \citenamefont {Baumgartner}, \citenamefont {Makk}, \citenamefont {Madsen},
  \citenamefont {Guzenko}, \citenamefont {Nyg\aa{}rd}, \citenamefont
  {Sch\"onenberger}, \citenamefont {Levy~Yeyati},\ and\ \citenamefont
  {Csonka}}]{Fulop_PRL2015}%
  \BibitemOpen
  \bibfield  {author} {\bibinfo {author} {\bibfnamefont {G.}~\bibnamefont
  {F\"ul\"op}}, \bibinfo {author} {\bibfnamefont {F.}~\bibnamefont
  {Dom\'{\i}nguez}}, \bibinfo {author} {\bibfnamefont {S.}~\bibnamefont
  {d'Hollosy}}, \bibinfo {author} {\bibfnamefont {A.}~\bibnamefont
  {Baumgartner}}, \bibinfo {author} {\bibfnamefont {P.}~\bibnamefont {Makk}},
  \bibinfo {author} {\bibfnamefont {M.~H.}\ \bibnamefont {Madsen}}, \bibinfo
  {author} {\bibfnamefont {V.~A.}\ \bibnamefont {Guzenko}}, \bibinfo {author}
  {\bibfnamefont {J.}~\bibnamefont {Nyg\aa{}rd}}, \bibinfo {author}
  {\bibfnamefont {C.}~\bibnamefont {Sch\"onenberger}}, \bibinfo {author}
  {\bibfnamefont {A.}~\bibnamefont {Levy~Yeyati}},\ and\ \bibinfo {author}
  {\bibfnamefont {S.}~\bibnamefont {Csonka}},\ }\href
  {https://doi.org/10.1103/PhysRevLett.115.227003} {\bibfield  {journal}
  {\bibinfo  {journal} {Phys. Rev. Lett.}\ }\textbf {\bibinfo {volume} {115}},\
  \bibinfo {pages} {227003} (\bibinfo {year} {2015})}\BibitemShut {NoStop}%
\bibitem [{\citenamefont {Liu}\ \emph {et~al.}(2022)\citenamefont {Liu},
  \citenamefont {Wang}, \citenamefont {Dvir},\ and\ \citenamefont
  {Wimmer}}]{Liu_arXiv2022}%
  \BibitemOpen
  \bibfield  {author} {\bibinfo {author} {\bibfnamefont {C.-X.}\ \bibnamefont
  {Liu}}, \bibinfo {author} {\bibfnamefont {G.}~\bibnamefont {Wang}}, \bibinfo
  {author} {\bibfnamefont {T.}~\bibnamefont {Dvir}},\ and\ \bibinfo {author}
  {\bibfnamefont {M.}~\bibnamefont {Wimmer}},\ }\href
  {https://arxiv.org/abs/2203.00107} {\bibfield  {journal} {\bibinfo  {journal}
  {arXiv:2203.00107}\ } (\bibinfo {year} {2022})}\BibitemShut {NoStop}%
\bibitem [{\citenamefont {Aksenov}\ \emph {et~al.}(2020)\citenamefont
  {Aksenov}, \citenamefont {Zlotnikov},\ and\ \citenamefont
  {Shustin}}]{Aksenov2020}%
  \BibitemOpen
  \bibfield  {author} {\bibinfo {author} {\bibfnamefont {S.~V.}\ \bibnamefont
  {Aksenov}}, \bibinfo {author} {\bibfnamefont {A.~O.}\ \bibnamefont
  {Zlotnikov}},\ and\ \bibinfo {author} {\bibfnamefont {M.~S.}\ \bibnamefont
  {Shustin}},\ }\href@noop {} {\bibfield  {journal} {\bibinfo  {journal} {Phys.
  Rev. B}\ }\textbf {\bibinfo {volume} {101}},\ \bibinfo {pages} {125431}
  (\bibinfo {year} {2020})}\BibitemShut {NoStop}%
\bibitem [{\citenamefont {Sedlmayr}\ and\ \citenamefont
  {Bena}(2015)}]{Sedlmayr2015}%
  \BibitemOpen
  \bibfield  {author} {\bibinfo {author} {\bibfnamefont {N.}~\bibnamefont
  {Sedlmayr}}\ and\ \bibinfo {author} {\bibfnamefont {C.}~\bibnamefont
  {Bena}},\ }\href@noop {} {\bibfield  {journal} {\bibinfo  {journal} {Phys.
  Rev. B}\ }\textbf {\bibinfo {volume} {92}},\ \bibinfo {pages} {115115}
  (\bibinfo {year} {2015})}\BibitemShut {NoStop}%
\bibitem [{\citenamefont {Sedlmayr}\ \emph {et~al.}(2016)\citenamefont
  {Sedlmayr}, \citenamefont {Aguiar-Hualde},\ and\ \citenamefont
  {Bena}}]{Sedlmayr2016}%
  \BibitemOpen
  \bibfield  {author} {\bibinfo {author} {\bibfnamefont {N.}~\bibnamefont
  {Sedlmayr}}, \bibinfo {author} {\bibfnamefont {J.~M.}\ \bibnamefont
  {Aguiar-Hualde}},\ and\ \bibinfo {author} {\bibfnamefont {C.}~\bibnamefont
  {Bena}},\ }\href@noop {} {\bibfield  {journal} {\bibinfo  {journal} {Phys.
  Rev. B}\ }\textbf {\bibinfo {volume} {93}},\ \bibinfo {pages} {155425}
  (\bibinfo {year} {2016})}\BibitemShut {NoStop}%
\bibitem [{\citenamefont {Dvir}\ \emph {et~al.}(2022)\citenamefont {Dvir},
  \citenamefont {Wang}, \citenamefont {van Loo}, \citenamefont {Liu},
  \citenamefont {Mazur}, \citenamefont {Bordin}, \citenamefont {ten Haaf},
  \citenamefont {Wang}, \citenamefont {van Driel}, \citenamefont {Zatelli},
  \citenamefont {Li}, \citenamefont {Malinowski}, \citenamefont {Gazibegovic},
  \citenamefont {Badawy}, \citenamefont {Bakkers}, \citenamefont {Wimmer},\
  and\ \citenamefont {Kouwenhoven}}]{Dvir_arXiv2022}%
  \BibitemOpen
  \bibfield  {author} {\bibinfo {author} {\bibfnamefont {T.}~\bibnamefont
  {Dvir}}, \bibinfo {author} {\bibfnamefont {G.}~\bibnamefont {Wang}}, \bibinfo
  {author} {\bibfnamefont {N.}~\bibnamefont {van Loo}}, \bibinfo {author}
  {\bibfnamefont {C.-X.}\ \bibnamefont {Liu}}, \bibinfo {author} {\bibfnamefont
  {G.~P.}\ \bibnamefont {Mazur}}, \bibinfo {author} {\bibfnamefont
  {A.}~\bibnamefont {Bordin}}, \bibinfo {author} {\bibfnamefont {S.~L.~D.}\
  \bibnamefont {ten Haaf}}, \bibinfo {author} {\bibfnamefont {J.-Y.}\
  \bibnamefont {Wang}}, \bibinfo {author} {\bibfnamefont {D.}~\bibnamefont {van
  Driel}}, \bibinfo {author} {\bibfnamefont {F.}~\bibnamefont {Zatelli}},
  \bibinfo {author} {\bibfnamefont {X.}~\bibnamefont {Li}}, \bibinfo {author}
  {\bibfnamefont {F.~K.}\ \bibnamefont {Malinowski}}, \bibinfo {author}
  {\bibfnamefont {S.}~\bibnamefont {Gazibegovic}}, \bibinfo {author}
  {\bibfnamefont {G.}~\bibnamefont {Badawy}}, \bibinfo {author} {\bibfnamefont
  {E.~P. A.~M.}\ \bibnamefont {Bakkers}}, \bibinfo {author} {\bibfnamefont
  {M.}~\bibnamefont {Wimmer}},\ and\ \bibinfo {author} {\bibfnamefont {L.~P.}\
  \bibnamefont {Kouwenhoven}},\ }\href {https://arxiv.org/abs/2206.08045}
  {\bibfield  {journal} {\bibinfo  {journal} {arXiv:2206.08045}\ } (\bibinfo
  {year} {2022})}\BibitemShut {NoStop}%
\bibitem [{\citenamefont {Brunetti}\ \emph {et~al.}(2013)\citenamefont
  {Brunetti}, \citenamefont {Zazunov}, \citenamefont {Kundu},\ and\
  \citenamefont {Egger}}]{PhysRevB.88.144515}%
  \BibitemOpen
  \bibfield  {author} {\bibinfo {author} {\bibfnamefont {A.}~\bibnamefont
  {Brunetti}}, \bibinfo {author} {\bibfnamefont {A.}~\bibnamefont {Zazunov}},
  \bibinfo {author} {\bibfnamefont {A.}~\bibnamefont {Kundu}},\ and\ \bibinfo
  {author} {\bibfnamefont {R.}~\bibnamefont {Egger}},\ }\href
  {https://doi.org/10.1103/PhysRevB.88.144515} {\bibfield  {journal} {\bibinfo
  {journal} {Phys. Rev. B}\ }\textbf {\bibinfo {volume} {88}},\ \bibinfo
  {pages} {144515} (\bibinfo {year} {2013})}\BibitemShut {NoStop}%
\bibitem [{\citenamefont {Wright}\ and\ \citenamefont
  {Veldhorst}(2013)}]{Wright_PRL2013}%
  \BibitemOpen
  \bibfield  {author} {\bibinfo {author} {\bibfnamefont {A.~R.}\ \bibnamefont
  {Wright}}\ and\ \bibinfo {author} {\bibfnamefont {M.}~\bibnamefont
  {Veldhorst}},\ }\href {https://doi.org/10.1103/PhysRevLett.111.096801}
  {\bibfield  {journal} {\bibinfo  {journal} {Phys. Rev. Lett.}\ }\textbf
  {\bibinfo {volume} {111}},\ \bibinfo {pages} {096801} (\bibinfo {year}
  {2013})}\BibitemShut {NoStop}%
\bibitem [{\citenamefont {O'Brien}\ \emph {et~al.}(2015)\citenamefont
  {O'Brien}, \citenamefont {Wright},\ and\ \citenamefont
  {Veldhorst}}]{OBrien2015}%
  \BibitemOpen
  \bibfield  {author} {\bibinfo {author} {\bibfnamefont {T.~E.}\ \bibnamefont
  {O'Brien}}, \bibinfo {author} {\bibfnamefont {A.~R.}\ \bibnamefont
  {Wright}},\ and\ \bibinfo {author} {\bibfnamefont {M.}~\bibnamefont
  {Veldhorst}},\ }\href@noop {} {\bibfield  {journal} {\bibinfo  {journal}
  {Phys. status solidi}\ }\textbf {\bibinfo {volume} {252}},\ \bibinfo {pages}
  {1731} (\bibinfo {year} {2015})}\BibitemShut {NoStop}%
\bibitem [{\citenamefont {Stepanenko}\ \emph {et~al.}(2012)\citenamefont
  {Stepanenko}, \citenamefont {Rudner}, \citenamefont {Halperin},\ and\
  \citenamefont {Loss}}]{Stepanenko2012}%
  \BibitemOpen
  \bibfield  {author} {\bibinfo {author} {\bibfnamefont {D.}~\bibnamefont
  {Stepanenko}}, \bibinfo {author} {\bibfnamefont {M.}~\bibnamefont {Rudner}},
  \bibinfo {author} {\bibfnamefont {B.~I.}\ \bibnamefont {Halperin}},\ and\
  \bibinfo {author} {\bibfnamefont {D.}~\bibnamefont {Loss}},\ }\href@noop {}
  {\bibfield  {journal} {\bibinfo  {journal} {Phys. Rev. B}\ }\textbf {\bibinfo
  {volume} {85}},\ \bibinfo {pages} {075416} (\bibinfo {year}
  {2012})}\BibitemShut {NoStop}%
\bibitem [{\citenamefont {Governale}\ \emph {et~al.}(2008)\citenamefont
  {Governale}, \citenamefont {Pala},\ and\ \citenamefont
  {K{\"o}nig}}]{Governale2008}%
  \BibitemOpen
  \bibfield  {author} {\bibinfo {author} {\bibfnamefont {M.}~\bibnamefont
  {Governale}}, \bibinfo {author} {\bibfnamefont {M.~G.}\ \bibnamefont
  {Pala}},\ and\ \bibinfo {author} {\bibfnamefont {J.}~\bibnamefont
  {K{\"o}nig}},\ }\href@noop {} {\bibfield  {journal} {\bibinfo  {journal}
  {Phys. Rev. B}\ }\textbf {\bibinfo {volume} {77}},\ \bibinfo {pages} {134513}
  (\bibinfo {year} {2008})}\BibitemShut {NoStop}%
\bibitem [{\citenamefont {Rozhkov}\ and\ \citenamefont
  {Arovas}(2000)}]{Rozhkov2000}%
  \BibitemOpen
  \bibfield  {author} {\bibinfo {author} {\bibfnamefont {A.~V.}\ \bibnamefont
  {Rozhkov}}\ and\ \bibinfo {author} {\bibfnamefont {D.~P.}\ \bibnamefont
  {Arovas}},\ }\href@noop {} {\bibfield  {journal} {\bibinfo  {journal} {Phys.
  Rev. B}\ }\textbf {\bibinfo {volume} {62}},\ \bibinfo {pages} {6687}
  (\bibinfo {year} {2000})}\BibitemShut {NoStop}%
\bibitem [{\citenamefont {Tanaka}\ \emph {et~al.}(2007)\citenamefont {Tanaka},
  \citenamefont {Oguri},\ and\ \citenamefont {Hewson}}]{Tanaka2007}%
  \BibitemOpen
  \bibfield  {author} {\bibinfo {author} {\bibfnamefont {Y.}~\bibnamefont
  {Tanaka}}, \bibinfo {author} {\bibfnamefont {A.}~\bibnamefont {Oguri}},\ and\
  \bibinfo {author} {\bibfnamefont {A.~C.}\ \bibnamefont {Hewson}},\
  }\href@noop {} {\bibfield  {journal} {\bibinfo  {journal} {New J. Phys.}\
  }\textbf {\bibinfo {volume} {9}},\ \bibinfo {pages} {115} (\bibinfo {year}
  {2007})}\BibitemShut {NoStop}%
\bibitem [{\citenamefont {Karrasch}\ \emph {et~al.}(2008)\citenamefont
  {Karrasch}, \citenamefont {Oguri},\ and\ \citenamefont
  {Meden}}]{Karrasch2008}%
  \BibitemOpen
  \bibfield  {author} {\bibinfo {author} {\bibfnamefont {C.}~\bibnamefont
  {Karrasch}}, \bibinfo {author} {\bibfnamefont {A.}~\bibnamefont {Oguri}},\
  and\ \bibinfo {author} {\bibfnamefont {V.}~\bibnamefont {Meden}},\
  }\href@noop {} {\bibfield  {journal} {\bibinfo  {journal} {Phys. Rev. B}\
  }\textbf {\bibinfo {volume} {77}},\ \bibinfo {pages} {024517} (\bibinfo
  {year} {2008})}\BibitemShut {NoStop}%
\bibitem [{\citenamefont {Wang}\ \emph {et~al.}(2022)\citenamefont {Wang},
  \citenamefont {Dvir}, \citenamefont {Mazur}, \citenamefont {Liu},
  \citenamefont {van Loo}, \citenamefont {ten Haaf}, \citenamefont {Bordin},
  \citenamefont {Gazibegovic}, \citenamefont {Badawy}, \citenamefont {Bakkers},
  \citenamefont {Wimmer},\ and\ \citenamefont {Kouwenhoven}}]{Wang2022}%
  \BibitemOpen
  \bibfield  {author} {\bibinfo {author} {\bibfnamefont {G.}~\bibnamefont
  {Wang}}, \bibinfo {author} {\bibfnamefont {T.}~\bibnamefont {Dvir}}, \bibinfo
  {author} {\bibfnamefont {G.~P.}\ \bibnamefont {Mazur}}, \bibinfo {author}
  {\bibfnamefont {C.-X.}\ \bibnamefont {Liu}}, \bibinfo {author} {\bibfnamefont
  {N.}~\bibnamefont {van Loo}}, \bibinfo {author} {\bibfnamefont {S.~L.~D.}\
  \bibnamefont {ten Haaf}}, \bibinfo {author} {\bibfnamefont {A.}~\bibnamefont
  {Bordin}}, \bibinfo {author} {\bibfnamefont {S.}~\bibnamefont {Gazibegovic}},
  \bibinfo {author} {\bibfnamefont {G.}~\bibnamefont {Badawy}}, \bibinfo
  {author} {\bibfnamefont {E.~P. A.~M.}\ \bibnamefont {Bakkers}}, \bibinfo
  {author} {\bibfnamefont {M.}~\bibnamefont {Wimmer}},\ and\ \bibinfo {author}
  {\bibfnamefont {L.~P.}\ \bibnamefont {Kouwenhoven}},\ }\href@noop {}
  {\bibfield  {journal} {\bibinfo  {journal} {arXiv:2205.03458}\ } (\bibinfo
  {year} {2022})}\BibitemShut {NoStop}%
\bibitem [{SI()}]{SI}%
  \BibitemOpen
  \href@noop {} {}\bibinfo {note} {See Supplementary Information in (URL
  included by the Editor) including Refs.
  \cite{HenrikBruus2004,Grove_NatCom2018,YU1965,Shiba1968,Rusinov1969}}\BibitemShut
  {NoStop}%
\bibitem [{\citenamefont {Kiršanskas}\ \emph {et~al.}(2017)\citenamefont
  {Kiršanskas}, \citenamefont {Pedersen}, \citenamefont {Karlström},
  \citenamefont {Leijnse},\ and\ \citenamefont {Wacker}}]{Kirsanskas_CPC2017}%
  \BibitemOpen
  \bibfield  {author} {\bibinfo {author} {\bibfnamefont {G.}~\bibnamefont
  {Kiršanskas}}, \bibinfo {author} {\bibfnamefont {J.~N.}\ \bibnamefont
  {Pedersen}}, \bibinfo {author} {\bibfnamefont {O.}~\bibnamefont
  {Karlström}}, \bibinfo {author} {\bibfnamefont {M.}~\bibnamefont
  {Leijnse}},\ and\ \bibinfo {author} {\bibfnamefont {A.}~\bibnamefont
  {Wacker}},\ }\href
  {https://doi.org/https://doi.org/10.1016/j.cpc.2017.07.024} {\bibfield
  {journal} {\bibinfo  {journal} {Comput. Phys. Commun.}\ }\textbf {\bibinfo
  {volume} {221}},\ \bibinfo {pages} {317} (\bibinfo {year}
  {2017})}\BibitemShut {NoStop}%
\bibitem [{\citenamefont {Pikulin}\ \emph {et~al.}(2021)\citenamefont
  {Pikulin}, \citenamefont {van Heck}, \citenamefont {Karzig}, \citenamefont
  {Martinez}, \citenamefont {Nijholt}, \citenamefont {Laeven}, \citenamefont
  {Winkler}, \citenamefont {Watson}, \citenamefont {Heedt}, \citenamefont
  {Temurhan}, \citenamefont {Svidenko}, \citenamefont {Lutchyn}, \citenamefont
  {Thomas}, \citenamefont {de~Lange}, \citenamefont {Casparis},\ and\
  \citenamefont {Nayak}}]{Pikulin2021}%
  \BibitemOpen
  \bibfield  {author} {\bibinfo {author} {\bibfnamefont {D.~I.}\ \bibnamefont
  {Pikulin}}, \bibinfo {author} {\bibfnamefont {B.}~\bibnamefont {van Heck}},
  \bibinfo {author} {\bibfnamefont {T.}~\bibnamefont {Karzig}}, \bibinfo
  {author} {\bibfnamefont {E.~A.}\ \bibnamefont {Martinez}}, \bibinfo {author}
  {\bibfnamefont {B.}~\bibnamefont {Nijholt}}, \bibinfo {author} {\bibfnamefont
  {T.}~\bibnamefont {Laeven}}, \bibinfo {author} {\bibfnamefont {G.~W.}\
  \bibnamefont {Winkler}}, \bibinfo {author} {\bibfnamefont {J.~D.}\
  \bibnamefont {Watson}}, \bibinfo {author} {\bibfnamefont {S.}~\bibnamefont
  {Heedt}}, \bibinfo {author} {\bibfnamefont {M.}~\bibnamefont {Temurhan}},
  \bibinfo {author} {\bibfnamefont {V.}~\bibnamefont {Svidenko}}, \bibinfo
  {author} {\bibfnamefont {R.~M.}\ \bibnamefont {Lutchyn}}, \bibinfo {author}
  {\bibfnamefont {M.}~\bibnamefont {Thomas}}, \bibinfo {author} {\bibfnamefont
  {G.}~\bibnamefont {de~Lange}}, \bibinfo {author} {\bibfnamefont
  {L.}~\bibnamefont {Casparis}},\ and\ \bibinfo {author} {\bibfnamefont
  {C.}~\bibnamefont {Nayak}},\ }\href@noop {} {\bibfield  {journal} {\bibinfo
  {journal} {arXiv:2103.12217v1}\ } (\bibinfo {year} {2021})}\BibitemShut
  {NoStop}%
\bibitem [{\citenamefont {Danon}\ \emph {et~al.}(2020)\citenamefont {Danon},
  \citenamefont {Hellenes}, \citenamefont {Hansen}, \citenamefont {Casparis},
  \citenamefont {Higginbotham},\ and\ \citenamefont {Flensberg}}]{Danon2020}%
  \BibitemOpen
  \bibfield  {author} {\bibinfo {author} {\bibfnamefont {J.}~\bibnamefont
  {Danon}}, \bibinfo {author} {\bibfnamefont {A.~B.}\ \bibnamefont {Hellenes}},
  \bibinfo {author} {\bibfnamefont {E.~B.}\ \bibnamefont {Hansen}}, \bibinfo
  {author} {\bibfnamefont {L.}~\bibnamefont {Casparis}}, \bibinfo {author}
  {\bibfnamefont {A.~P.}\ \bibnamefont {Higginbotham}},\ and\ \bibinfo {author}
  {\bibfnamefont {K.}~\bibnamefont {Flensberg}},\ }\href@noop {} {\bibfield
  {journal} {\bibinfo  {journal} {Phys. Rev. Lett.}\ }\textbf {\bibinfo
  {volume} {124}},\ \bibinfo {pages} {036801} (\bibinfo {year}
  {2020})}\BibitemShut {NoStop}%
\bibitem [{\citenamefont {Pöschl}\ \emph {et~al.}(2022)\citenamefont
  {Pöschl}, \citenamefont {Danilenko}, \citenamefont {Sabonis}, \citenamefont
  {Kristjuhan}, \citenamefont {Lindemann}, \citenamefont {Thomas},
  \citenamefont {Manfra},\ and\ \citenamefont {Marcus}}]{Poeschl2022}%
  \BibitemOpen
  \bibfield  {author} {\bibinfo {author} {\bibfnamefont {A.}~\bibnamefont
  {Pöschl}}, \bibinfo {author} {\bibfnamefont {A.}~\bibnamefont {Danilenko}},
  \bibinfo {author} {\bibfnamefont {D.}~\bibnamefont {Sabonis}}, \bibinfo
  {author} {\bibfnamefont {K.}~\bibnamefont {Kristjuhan}}, \bibinfo {author}
  {\bibfnamefont {T.}~\bibnamefont {Lindemann}}, \bibinfo {author}
  {\bibfnamefont {C.}~\bibnamefont {Thomas}}, \bibinfo {author} {\bibfnamefont
  {M.~J.}\ \bibnamefont {Manfra}},\ and\ \bibinfo {author} {\bibfnamefont
  {C.~M.}\ \bibnamefont {Marcus}},\ }\href@noop {} {\bibfield  {journal}
  {\bibinfo  {journal} {arXiv:2204.02430v1}\ } (\bibinfo {year}
  {2022})}\BibitemShut {NoStop}%
\bibitem [{\citenamefont {Maiani}\ \emph {et~al.}(2022)\citenamefont {Maiani},
  \citenamefont {Geier},\ and\ \citenamefont {Flensberg}}]{Maiani2022}%
  \BibitemOpen
  \bibfield  {author} {\bibinfo {author} {\bibfnamefont {A.}~\bibnamefont
  {Maiani}}, \bibinfo {author} {\bibfnamefont {M.}~\bibnamefont {Geier}},\ and\
  \bibinfo {author} {\bibfnamefont {K.}~\bibnamefont {Flensberg}},\ }\href@noop
  {} {\bibfield  {journal} {\bibinfo  {journal} {arXiv:2205.11193v2}\ }
  (\bibinfo {year} {2022})}\BibitemShut {NoStop}%
\bibitem [{\citenamefont {Liu}\ \emph {et~al.}(2015{\natexlab{a}})\citenamefont
  {Liu}, \citenamefont {Cheng},\ and\ \citenamefont {Lutchyn}}]{Liu2015}%
  \BibitemOpen
  \bibfield  {author} {\bibinfo {author} {\bibfnamefont {D.~E.}\ \bibnamefont
  {Liu}}, \bibinfo {author} {\bibfnamefont {M.}~\bibnamefont {Cheng}},\ and\
  \bibinfo {author} {\bibfnamefont {R.~M.}\ \bibnamefont {Lutchyn}},\ }\href
  {https://doi.org/10.1103/physrevb.91.081405} {\bibfield  {journal} {\bibinfo
  {journal} {Phys. Rev. B}\ }\textbf {\bibinfo {volume} {91}},\ \bibinfo
  {pages} {081405(R)} (\bibinfo {year} {2015}{\natexlab{a}})}\BibitemShut
  {NoStop}%
\bibitem [{\citenamefont {Liu}\ \emph {et~al.}(2015{\natexlab{b}})\citenamefont
  {Liu}, \citenamefont {Levchenko},\ and\ \citenamefont {Lutchyn}}]{Liu2015a}%
  \BibitemOpen
  \bibfield  {author} {\bibinfo {author} {\bibfnamefont {D.~E.}\ \bibnamefont
  {Liu}}, \bibinfo {author} {\bibfnamefont {A.}~\bibnamefont {Levchenko}},\
  and\ \bibinfo {author} {\bibfnamefont {R.~M.}\ \bibnamefont {Lutchyn}},\
  }\href {https://doi.org/10.1103/physrevb.92.205422} {\bibfield  {journal}
  {\bibinfo  {journal} {Phys. Rev. B}\ }\textbf {\bibinfo {volume} {92}},\
  \bibinfo {pages} {205422} (\bibinfo {year} {2015}{\natexlab{b}})}\BibitemShut
  {NoStop}%
\bibitem [{\citenamefont {Smirnov}(2019)}]{Smirnov2019}%
  \BibitemOpen
  \bibfield  {author} {\bibinfo {author} {\bibfnamefont {S.}~\bibnamefont
  {Smirnov}},\ }\href {https://doi.org/10.1103/physrevb.99.165427} {\bibfield
  {journal} {\bibinfo  {journal} {Phys. Rev. B}\ }\textbf {\bibinfo {volume}
  {99}},\ \bibinfo {pages} {165427} (\bibinfo {year} {2019})}\BibitemShut
  {NoStop}%
\bibitem [{\citenamefont {Feng}\ and\ \citenamefont {Zhang}(2022)}]{Feng2022}%
  \BibitemOpen
  \bibfield  {author} {\bibinfo {author} {\bibfnamefont {G.-H.}\ \bibnamefont
  {Feng}}\ and\ \bibinfo {author} {\bibfnamefont {H.-H.}\ \bibnamefont
  {Zhang}},\ }\href {https://doi.org/10.1103/physrevb.105.035148} {\bibfield
  {journal} {\bibinfo  {journal} {Phys. Rev. B}\ }\textbf {\bibinfo {volume}
  {105}},\ \bibinfo {pages} {035148} (\bibinfo {year} {2022})}\BibitemShut
  {NoStop}%
\bibitem [{\citenamefont {Smirnov}(2015)}]{Smirnov2015}%
  \BibitemOpen
  \bibfield  {author} {\bibinfo {author} {\bibfnamefont {S.}~\bibnamefont
  {Smirnov}},\ }\href {https://doi.org/10.1103/physrevb.92.195312} {\bibfield
  {journal} {\bibinfo  {journal} {Phys. Rev. B}\ }\textbf {\bibinfo {volume}
  {92}},\ \bibinfo {pages} {195312} (\bibinfo {year} {2015})}\BibitemShut
  {NoStop}%
\bibitem [{\citenamefont {Sela}\ \emph {et~al.}(2019)\citenamefont {Sela},
  \citenamefont {Oreg}, \citenamefont {Plugge}, \citenamefont {Hartman},
  \citenamefont {L{\"u}scher},\ and\ \citenamefont {Folk}}]{Sela2019}%
  \BibitemOpen
  \bibfield  {author} {\bibinfo {author} {\bibfnamefont {E.}~\bibnamefont
  {Sela}}, \bibinfo {author} {\bibfnamefont {Y.}~\bibnamefont {Oreg}}, \bibinfo
  {author} {\bibfnamefont {S.}~\bibnamefont {Plugge}}, \bibinfo {author}
  {\bibfnamefont {N.}~\bibnamefont {Hartman}}, \bibinfo {author} {\bibfnamefont
  {S.}~\bibnamefont {L{\"u}scher}},\ and\ \bibinfo {author} {\bibfnamefont
  {J.}~\bibnamefont {Folk}},\ }\href
  {https://doi.org/10.1103/physrevlett.123.147702} {\bibfield  {journal}
  {\bibinfo  {journal} {Phys. Rev. Lett.}\ }\textbf {\bibinfo {volume} {123}},\
  \bibinfo {pages} {147702} (\bibinfo {year} {2019})}\BibitemShut {NoStop}%
\bibitem [{\citenamefont {Smirnov}(2021)}]{Smirnov2021}%
  \BibitemOpen
  \bibfield  {author} {\bibinfo {author} {\bibfnamefont {S.}~\bibnamefont
  {Smirnov}},\ }\href {https://doi.org/10.1103/physrevb.103.075440} {\bibfield
  {journal} {\bibinfo  {journal} {Phys. Rev. B}\ }\textbf {\bibinfo {volume}
  {103}},\ \bibinfo {pages} {075440} (\bibinfo {year} {2021})}\BibitemShut
  {NoStop}%
\bibitem [{\citenamefont {Han}\ \emph {et~al.}(2022)\citenamefont {Han},
  \citenamefont {Iftikhar}, \citenamefont {Kleeorin}, \citenamefont {Anthore},
  \citenamefont {Pierre}, \citenamefont {Meir}, \citenamefont {Mitchell},\ and\
  \citenamefont {Sela}}]{Han2022}%
  \BibitemOpen
  \bibfield  {author} {\bibinfo {author} {\bibfnamefont {C.}~\bibnamefont
  {Han}}, \bibinfo {author} {\bibfnamefont {Z.}~\bibnamefont {Iftikhar}},
  \bibinfo {author} {\bibfnamefont {Y.}~\bibnamefont {Kleeorin}}, \bibinfo
  {author} {\bibfnamefont {A.}~\bibnamefont {Anthore}}, \bibinfo {author}
  {\bibfnamefont {F.}~\bibnamefont {Pierre}}, \bibinfo {author} {\bibfnamefont
  {Y.}~\bibnamefont {Meir}}, \bibinfo {author} {\bibfnamefont {A.~K.}\
  \bibnamefont {Mitchell}},\ and\ \bibinfo {author} {\bibfnamefont
  {E.}~\bibnamefont {Sela}},\ }\href
  {https://doi.org/10.1103/physrevlett.128.146803} {\bibfield  {journal}
  {\bibinfo  {journal} {Phys. Rev. Lett.}\ }\textbf {\bibinfo {volume} {128}},\
  \bibinfo {pages} {146803} (\bibinfo {year} {2022})}\BibitemShut {NoStop}%
\bibitem [{\citenamefont {Aasen}\ \emph {et~al.}(2016)\citenamefont {Aasen},
  \citenamefont {Hell}, \citenamefont {Mishmash}, \citenamefont {Higginbotham},
  \citenamefont {Danon}, \citenamefont {Leijnse}, \citenamefont {Jespersen},
  \citenamefont {Folk}, \citenamefont {Marcus}, \citenamefont {Flensberg},\
  and\ \citenamefont {Alicea}}]{Aasen_PRX2016}%
  \BibitemOpen
  \bibfield  {author} {\bibinfo {author} {\bibfnamefont {D.}~\bibnamefont
  {Aasen}}, \bibinfo {author} {\bibfnamefont {M.}~\bibnamefont {Hell}},
  \bibinfo {author} {\bibfnamefont {R.~V.}\ \bibnamefont {Mishmash}}, \bibinfo
  {author} {\bibfnamefont {A.}~\bibnamefont {Higginbotham}}, \bibinfo {author}
  {\bibfnamefont {J.}~\bibnamefont {Danon}}, \bibinfo {author} {\bibfnamefont
  {M.}~\bibnamefont {Leijnse}}, \bibinfo {author} {\bibfnamefont {T.~S.}\
  \bibnamefont {Jespersen}}, \bibinfo {author} {\bibfnamefont {J.~A.}\
  \bibnamefont {Folk}}, \bibinfo {author} {\bibfnamefont {C.~M.}\ \bibnamefont
  {Marcus}}, \bibinfo {author} {\bibfnamefont {K.}~\bibnamefont {Flensberg}},\
  and\ \bibinfo {author} {\bibfnamefont {J.}~\bibnamefont {Alicea}},\ }\href
  {https://doi.org/10.1103/PhysRevX.6.031016} {\bibfield  {journal} {\bibinfo
  {journal} {Phys. Rev. X}\ }\textbf {\bibinfo {volume} {6}},\ \bibinfo {pages}
  {031016} (\bibinfo {year} {2016})}\BibitemShut {NoStop}%
\bibitem [{\citenamefont {Souto}\ and\ \citenamefont
  {Leijnse}(2022)}]{Souto_SciPost2022}%
  \BibitemOpen
  \bibfield  {author} {\bibinfo {author} {\bibfnamefont {R.~S.}\ \bibnamefont
  {Souto}}\ and\ \bibinfo {author} {\bibfnamefont {M.}~\bibnamefont
  {Leijnse}},\ }\href {https://doi.org/10.21468/SciPostPhys.12.5.161}
  {\bibfield  {journal} {\bibinfo  {journal} {SciPost Phys.}\ }\textbf
  {\bibinfo {volume} {12}},\ \bibinfo {pages} {161} (\bibinfo {year}
  {2022})}\BibitemShut {NoStop}%
\bibitem [{\citenamefont {Bonderson}\ \emph {et~al.}(2008)\citenamefont
  {Bonderson}, \citenamefont {Freedman},\ and\ \citenamefont
  {Nayak}}]{Bonderson_PRL2008}%
  \BibitemOpen
  \bibfield  {author} {\bibinfo {author} {\bibfnamefont {P.}~\bibnamefont
  {Bonderson}}, \bibinfo {author} {\bibfnamefont {M.}~\bibnamefont
  {Freedman}},\ and\ \bibinfo {author} {\bibfnamefont {C.}~\bibnamefont
  {Nayak}},\ }\href {https://doi.org/10.1103/PhysRevLett.101.010501} {\bibfield
   {journal} {\bibinfo  {journal} {Phys. Rev. Lett.}\ }\textbf {\bibinfo
  {volume} {101}},\ \bibinfo {pages} {010501} (\bibinfo {year}
  {2008})}\BibitemShut {NoStop}%
\bibitem [{\citenamefont {Flensberg}(2011)}]{Flensberg_PRL2011}%
  \BibitemOpen
  \bibfield  {author} {\bibinfo {author} {\bibfnamefont {K.}~\bibnamefont
  {Flensberg}},\ }\href {https://doi.org/10.1103/PhysRevLett.106.090503}
  {\bibfield  {journal} {\bibinfo  {journal} {Phys. Rev. Lett.}\ }\textbf
  {\bibinfo {volume} {106}},\ \bibinfo {pages} {090503} (\bibinfo {year}
  {2011})}\BibitemShut {NoStop}%
\bibitem [{\citenamefont {van Heck}\ \emph {et~al.}(2012)\citenamefont {van
  Heck}, \citenamefont {Akhmerov}, \citenamefont {Hassler}, \citenamefont
  {Burrello},\ and\ \citenamefont {Beenakker}}]{van_Heck_NJP2012}%
  \BibitemOpen
  \bibfield  {author} {\bibinfo {author} {\bibfnamefont {B.}~\bibnamefont {van
  Heck}}, \bibinfo {author} {\bibfnamefont {A.~R.}\ \bibnamefont {Akhmerov}},
  \bibinfo {author} {\bibfnamefont {F.}~\bibnamefont {Hassler}}, \bibinfo
  {author} {\bibfnamefont {M.}~\bibnamefont {Burrello}},\ and\ \bibinfo
  {author} {\bibfnamefont {C.~W.~J.}\ \bibnamefont {Beenakker}},\ }\href
  {https://doi.org/10.1088/1367-2630/14/3/035019} {\bibfield  {journal}
  {\bibinfo  {journal} {New J. Phys.}\ }\textbf {\bibinfo {volume} {14}},\
  \bibinfo {pages} {035019} (\bibinfo {year} {2012})}\BibitemShut {NoStop}%
\bibitem [{\citenamefont {H.~Bruus}(2004)}]{HenrikBruus2004}%
  \BibitemOpen
  \bibfield  {author} {\bibinfo {author} {\bibfnamefont {K.~F.}\ \bibnamefont
  {H.~Bruus}},\ }\href@noop {} {\emph {\bibinfo {title} {Many-Body Quantum
  Theory in Condensed Matter Physics: An Introduction}}}\ (\bibinfo
  {publisher} {Oxford University Press},\ \bibinfo {year} {2004})\BibitemShut
  {NoStop}%
\bibitem [{\citenamefont {Grove-Rasmussen}\ \emph {et~al.}(2018)\citenamefont
  {Grove-Rasmussen}, \citenamefont {Steffensen}, \citenamefont {Jellinggaard},
  \citenamefont {Madsen}, \citenamefont {{\v{Z}}itko}, \citenamefont {Paaske},\
  and\ \citenamefont {Nyg{\aa}rd}}]{Grove_NatCom2018}%
  \BibitemOpen
  \bibfield  {author} {\bibinfo {author} {\bibfnamefont {K.}~\bibnamefont
  {Grove-Rasmussen}}, \bibinfo {author} {\bibfnamefont {G.}~\bibnamefont
  {Steffensen}}, \bibinfo {author} {\bibfnamefont {A.}~\bibnamefont
  {Jellinggaard}}, \bibinfo {author} {\bibfnamefont {M.~H.}\ \bibnamefont
  {Madsen}}, \bibinfo {author} {\bibfnamefont {R.}~\bibnamefont {{\v{Z}}itko}},
  \bibinfo {author} {\bibfnamefont {J.}~\bibnamefont {Paaske}},\ and\ \bibinfo
  {author} {\bibfnamefont {J.}~\bibnamefont {Nyg{\aa}rd}},\ }\href
  {https://doi.org/10.1038/s41467-018-04683-x} {\bibfield  {journal} {\bibinfo
  {journal} {Nature Commun.}\ }\textbf {\bibinfo {volume} {9}},\ \bibinfo
  {pages} {2376} (\bibinfo {year} {2018})}\BibitemShut {NoStop}%
\bibitem [{\citenamefont {Yu}(1965)}]{YU1965}%
  \BibitemOpen
  \bibfield  {author} {\bibinfo {author} {\bibfnamefont {L.}~\bibnamefont
  {Yu}},\ }\href {https://doi.org/10.7498/aps.21.75} {\bibfield  {journal}
  {\bibinfo  {journal} {Acta Phys. Sin.}\ }\textbf {\bibinfo {volume} {21}},\
  \bibinfo {pages} {75} (\bibinfo {year} {1965})}\BibitemShut {NoStop}%
\bibitem [{\citenamefont {Shiba}(1968)}]{Shiba1968}%
  \BibitemOpen
  \bibfield  {author} {\bibinfo {author} {\bibfnamefont {H.}~\bibnamefont
  {Shiba}},\ }\href@noop {} {\bibfield  {journal} {\bibinfo  {journal} {Prog.
  Theor. Phys.}\ }\textbf {\bibinfo {volume} {40}},\ \bibinfo {pages} {435}
  (\bibinfo {year} {1968})}\BibitemShut {NoStop}%
\bibitem [{\citenamefont {Rusinov}(1969)}]{Rusinov1969}%
  \BibitemOpen
  \bibfield  {author} {\bibinfo {author} {\bibfnamefont {A.~I.}\ \bibnamefont
  {Rusinov}},\ }\href@noop {} {\bibfield  {journal} {\bibinfo  {journal} {JETP
  Lett.}\ }\textbf {\bibinfo {volume} {9}},\ \bibinfo {pages} {85} (\bibinfo
  {year} {1969})}\BibitemShut {NoStop}%
\end{thebibliography}%


%apsrev4-2.bst 2019-01-14 (MD) hand-edited version of apsrev4-1.bst
%Control: key (0)
%Control: author (8) initials jnrlst
%Control: editor formatted (1) identically to author
%Control: production of article title (0) allowed
%Control: page (0) single
%Control: year (1) truncated
%Control: production of eprint (0) enabled
\begin{thebibliography}{5}%
\makeatletter
\providecommand \@ifxundefined [1]{%
 \@ifx{#1\undefined}
}%
\providecommand \@ifnum [1]{%
 \ifnum #1\expandafter \@firstoftwo
 \else \expandafter \@secondoftwo
 \fi
}%
\providecommand \@ifx [1]{%
 \ifx #1\expandafter \@firstoftwo
 \else \expandafter \@secondoftwo
 \fi
}%
\providecommand \natexlab [1]{#1}%
\providecommand \enquote  [1]{``#1''}%
\providecommand \bibnamefont  [1]{#1}%
\providecommand \bibfnamefont [1]{#1}%
\providecommand \citenamefont [1]{#1}%
\providecommand \href@noop [0]{\@secondoftwo}%
\providecommand \href [0]{\begingroup \@sanitize@url \@href}%
\providecommand \@href[1]{\@@startlink{#1}\@@href}%
\providecommand \@@href[1]{\endgroup#1\@@endlink}%
\providecommand \@sanitize@url [0]{\catcode `\\12\catcode `\$12\catcode
  `\&12\catcode `\#12\catcode `\^12\catcode `\_12\catcode `\%12\relax}%
\providecommand \@@startlink[1]{}%
\providecommand \@@endlink[0]{}%
\providecommand \url  [0]{\begingroup\@sanitize@url \@url }%
\providecommand \@url [1]{\endgroup\@href {#1}{\urlprefix }}%
\providecommand \urlprefix  [0]{URL }%
\providecommand \Eprint [0]{\href }%
\providecommand \doibase [0]{https://doi.org/}%
\providecommand \selectlanguage [0]{\@gobble}%
\providecommand \bibinfo  [0]{\@secondoftwo}%
\providecommand \bibfield  [0]{\@secondoftwo}%
\providecommand \translation [1]{[#1]}%
\providecommand \BibitemOpen [0]{}%
\providecommand \bibitemStop [0]{}%
\providecommand \bibitemNoStop [0]{.\EOS\space}%
\providecommand \EOS [0]{\spacefactor3000\relax}%
\providecommand \BibitemShut  [1]{\csname bibitem#1\endcsname}%
\let\auto@bib@innerbib\@empty
%</preamble>
\bibitem [{\citenamefont {Kiršanskas}\ \emph {et~al.}(2017)\citenamefont
  {Kiršanskas}, \citenamefont {Pedersen}, \citenamefont {Karlström},
  \citenamefont {Leijnse},\ and\ \citenamefont {Wacker}}]{Kirsanskas_CPC2017}%
  \BibitemOpen
  \bibfield  {author} {\bibinfo {author} {\bibfnamefont {G.}~\bibnamefont
  {Kiršanskas}}, \bibinfo {author} {\bibfnamefont {J.~N.}\ \bibnamefont
  {Pedersen}}, \bibinfo {author} {\bibfnamefont {O.}~\bibnamefont
  {Karlström}}, \bibinfo {author} {\bibfnamefont {M.}~\bibnamefont
  {Leijnse}},\ and\ \bibinfo {author} {\bibfnamefont {A.}~\bibnamefont
  {Wacker}},\ }\href
  {https://doi.org/https://doi.org/10.1016/j.cpc.2017.07.024} {\bibfield
  {journal} {\bibinfo  {journal} {Comput. Phys. Commun.}\ }\textbf {\bibinfo
  {volume} {221}},\ \bibinfo {pages} {317} (\bibinfo {year}
  {2017})}\BibitemShut {NoStop}%
\bibitem [{\citenamefont {H.~Bruus}(2004)}]{HenrikBruus2004}%
  \BibitemOpen
  \bibfield  {author} {\bibinfo {author} {\bibfnamefont {K.~F.}\ \bibnamefont
  {H.~Bruus}},\ }\href@noop {} {\emph {\bibinfo {title} {Many-Body Quantum
  Theory in Condensed Matter Physics: An Introduction}}}\ (\bibinfo
  {publisher} {Oxford University Press},\ \bibinfo {year} {2004})\BibitemShut
  {NoStop}%
\bibitem [{\citenamefont {Yu}(1965)}]{YU1965}%
  \BibitemOpen
  \bibfield  {author} {\bibinfo {author} {\bibfnamefont {L.}~\bibnamefont
  {Yu}},\ }\href {https://doi.org/10.7498/aps.21.75} {\bibfield  {journal}
  {\bibinfo  {journal} {Acta Phys. Sin.}\ }\textbf {\bibinfo {volume} {21}},\
  \bibinfo {pages} {75} (\bibinfo {year} {1965})}\BibitemShut {NoStop}%
\bibitem [{\citenamefont {Shiba}(1968)}]{Shiba1968}%
  \BibitemOpen
  \bibfield  {author} {\bibinfo {author} {\bibfnamefont {H.}~\bibnamefont
  {Shiba}},\ }\href@noop {} {\bibfield  {journal} {\bibinfo  {journal} {Prog.
  Theor. Phys.}\ }\textbf {\bibinfo {volume} {40}},\ \bibinfo {pages} {435}
  (\bibinfo {year} {1968})}\BibitemShut {NoStop}%
\bibitem [{\citenamefont {Rusinov}(1969)}]{Rusinov1969}%
  \BibitemOpen
  \bibfield  {author} {\bibinfo {author} {\bibfnamefont {A.~I.}\ \bibnamefont
  {Rusinov}},\ }\href@noop {} {\bibfield  {journal} {\bibinfo  {journal} {JETP
  Lett.}\ }\textbf {\bibinfo {volume} {9}},\ \bibinfo {pages} {85} (\bibinfo
  {year} {1969})}\BibitemShut {NoStop}%
\end{thebibliography}%

\end{document}

\setstcolor{blue}

\preprint{APS/123-QED}

\title{Supplemental Information to ``Creating and detecting poor man’s Majorana bound states
in interacting quantum dots''}%

\author{Athanasios Tsintzis}
\affiliation{%
Division of Solid State
Physics and NanoLund, Lund University, S-221 00 Lund, Sweden
}%
\author{Rub\'en Seoane Souto}%
\affiliation{%
Division of Solid State
Physics and NanoLund, Lund University, S-221 00 Lund, Sweden
}%
\affiliation{%
Center for Quantum Devices, Niels Bohr Institute, University of
Copenhagen, DK-2100 Copenhagen, Denmark
}%
\author{Martin Leijnse}
\affiliation{%
Division of Solid State
Physics and NanoLund, Lund University, S-221 00 Lund, Sweden
}%
\affiliation{%
Center for Quantum Devices, Niels Bohr Institute, University of
Copenhagen, DK-2100 Copenhagen, Denmark
}%

%\date{\today}% It is always \today, today,
             %  but any date may be explicitly specified

\maketitle

%\begin{widetext}
%In the Supplemental Material, we provide...
%\begin{itemize}
%    \item Rate equations 
%\end{itemize}
%\end{widetext}

\setcounter{figure}{0}
\setcounter{equation}{0}
%\twocolumngrid
\renewcommand\thefigure{S\arabic{figure}}
\renewcommand\theequation{S\arabic{equation}}

%%%%%%% Begin of the supplementary here
{\it Transport theory.}
Here, we provide the details of the transport theory used in the main text~\cite{Kirsanskas_CPC2017}, which is based on solving a rate equation for the reduced density matrix of the QD system. The rate equation neglects second and higher order processes in the tunnel coupling $\Gamma$ between a QD and the leads, which is valid in the regime $\Gamma \ll T$ ($e=\hbar=k_B=1$). In addition, it assumes the reduced density matrix of the QD system to be diagonal. This is a valid approximation when there are only (quasi-)degeneracies between eigenstates of the QD system that differ by a quantum number that is conserved by the total Hamiltonian, including the leads. In our case, the only such quantum number is the parity of the electron number. Here we motivate the formalism and present a form of the equations that does not rely on charge being a good quantum number in the system.

The full Hamiltonian of the system presented in Fig.~1(a) of the main text is:
\begin{equation}
    H = H_{QDs} + \sum_r H_r + H_T,
\end{equation}
where $H_{QDs}$ is given in Eq.~(1) in the main text and can be written in diagonalized form as $H_{QDs} =  \sum_a E_a \ket{a} \bra{a}$, with many-body eigenenergies $E_a$ and eigenstates $\ket{a}$ of definite parity. Note that charge is not a good quantum number because $H_{QDs}$ includes superconducting proximity effect on QD $C$. $\sum_r H_r$ is the sum of the normal (N) leads Hamiltonians ($r = L,R$) and $H_T$ contains the couplings between the N leads and the QDs.  $H_r$ and $H_T$ are given by:
\begin{align}
H_r =& \sum_{  k \sigma} \varepsilon_{r k \sigma } c^\dagger_{r k \sigma } c_{r k \sigma }, \\
H_T =& \sum_{r k \sigma  a} \Big[  T^{aa'}_{r \sigma +} \ket{a} \bra{a'} c_{r k \sigma} - T^{aa'}_{r \sigma -} \ket{a} \bra{a'} c^{\dagger}_{r k \sigma} \Big].
\end{align}
The operator $c^\dagger_{r k \sigma }$ creates an electron with spin $\sigma$ in momentum state $k$ in lead $r$ and $\varepsilon_{r k \sigma}$ is the corresponding electron energy. The transition matrix elements from QD state $\ket{a'}$ to $\ket{a}$, $T^{aa'}_{r \sigma + (-)}$, are given by the expressions:
\begin{align}
T^{aa'}_{r \sigma +} &= \sum_{j \neq C} \lambda_{r j \sigma} \bra{a} d^\dagger_{j \sigma} \ket{a'}, \\ T^{aa'}_{r \sigma -} &= \sum_{j \neq C} \lambda^*_{r j \sigma} \bra{a} d_{j \sigma} \ket{a'} = (T^{a'a}_{r \sigma +})^*,
\end{align}
where $\lambda_{r j \sigma}$ parametrizes the tunnel coupling between QD $j=L,R$ and lead $r$. The tunnel couplings can be calculated using Fermi's golden rule \cite{HenrikBruus2004}:
\begin{equation}
\Gamma^{aa'}_{r \sigma +(-)} = 2 \pi \rho_{r} |T^{aa'}_{r \sigma +(-)}|^2,
\end{equation}
where the density of states $\rho_{r}$ is taken to be energy-independent. The tunneling rates for jumping into and out of the QD are given by
\begin{align}
W^{aa'}_{\sigma +} &= \sum_r \Gamma^{aa'}_{r \sigma +} f_r(E_a - E_{a'}) = \sum_r W^{aa'}_{r \sigma +}, \\
W^{aa'}_{\sigma -} &= \sum_r \Gamma^{aa'}_{r \sigma -} \big[ 1- f_r(E_{a'}  -  E_a )\big] = \sum_r W^{aa'}_{r \sigma -},
\end{align}
where $f_r(x)$ is the Fermi function. The diagonal elements of the QD system's reduced density matrix are the occupation probabilities $P_a$ of the eigenstates $\ket{a}$, and their time dependence is given by:
\begin{equation}
\label{occ_prop_t_dep}
\dot{P}_a = \sum_{\sigma a'} \big[  W^{aa'}_{\sigma +} + W^{aa'}_{\sigma -}  \big] P_{a'} - 
\sum_{\sigma a'} \big[  W^{a'a}_{\sigma +} + W^{a'a}_{\sigma -}  \big] P_{a}.
\end{equation}
Note that the system can transition from state $\ket{a'}$ to state $\ket{a}$ both by adding ($+$) and removing ($-$) an electron to/from the QDs because charge is not a good quantum number. The set of Eqs.~(\ref{occ_prop_t_dep}) is solved for the steady state, $\dot{P}_a = 0$, together with the normalization condition $\sum_a P_a = 1$. The particle current flowing out of lead $r$ can be expressed by counting particles tunnelling into the QD system minus particles tunnelling out of the QD system, weighted by the probabilities of the initial states:
\begin{equation}
I_{r} = \sum_{\sigma a a'} (W^{aa'}_{r \sigma +} - W^{aa'}_{r \sigma -} )P_{a'}.
\end{equation}

\begin{figure*}
	\includegraphics[width=0.32\textwidth,trim={0cm 0cm 0cm 0cm},clip]{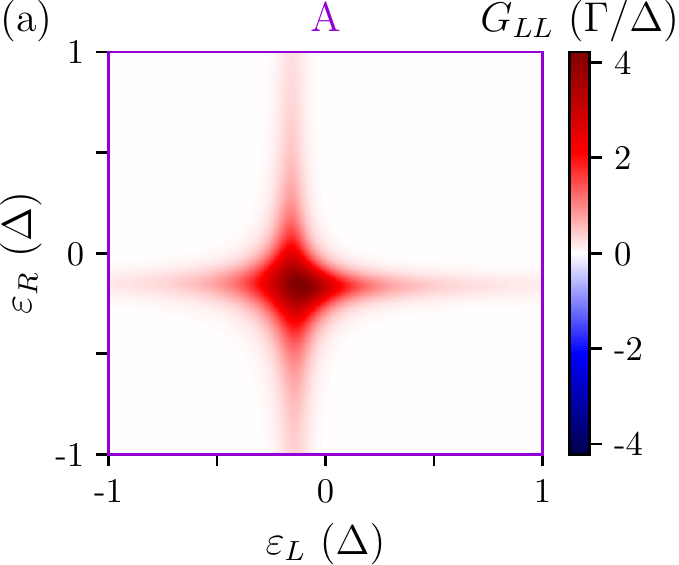}
	\includegraphics[width=0.32\textwidth,trim={0cm 0cm 0cm 0cm},clip]{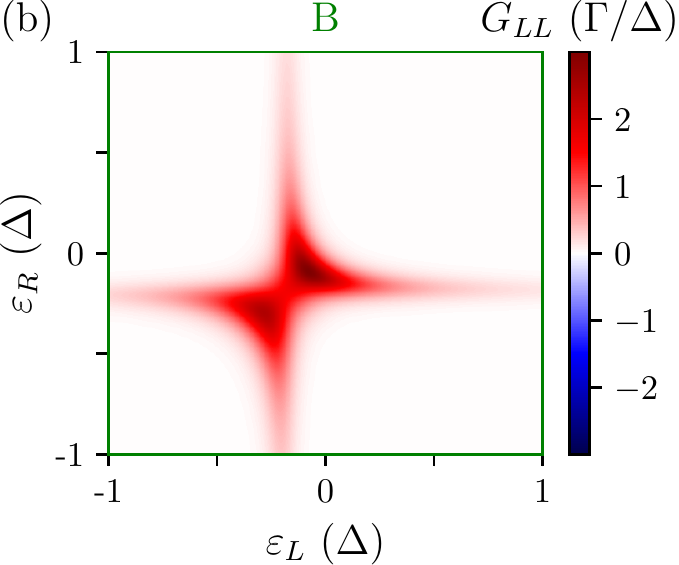}
	\includegraphics[width=0.32\textwidth,trim={0cm 0cm 0cm 0cm},clip]{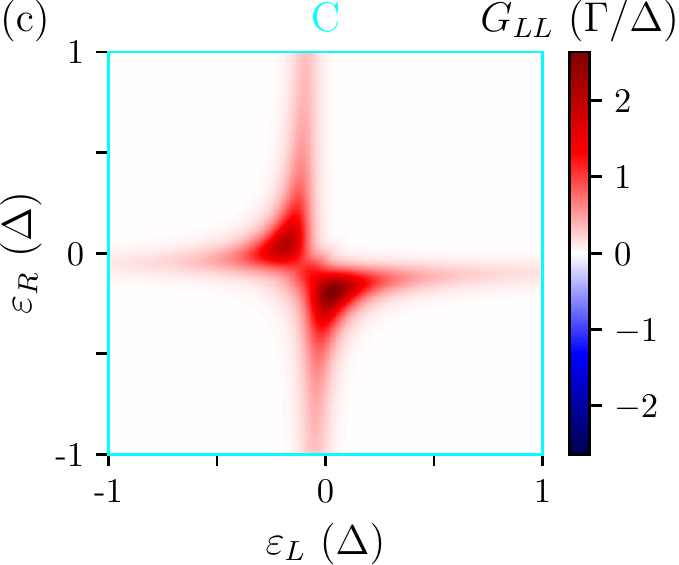}
	\includegraphics[width=0.47\textwidth,trim={0cm 0cm 0cm 0cm},clip]{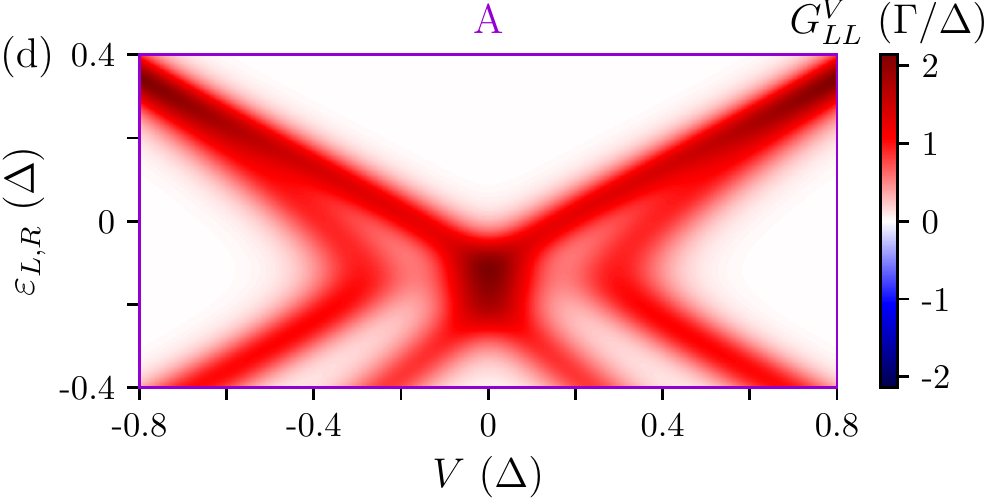}
	\includegraphics[width=0.47\textwidth,trim={0cm 0cm 0cm 0cm},clip]{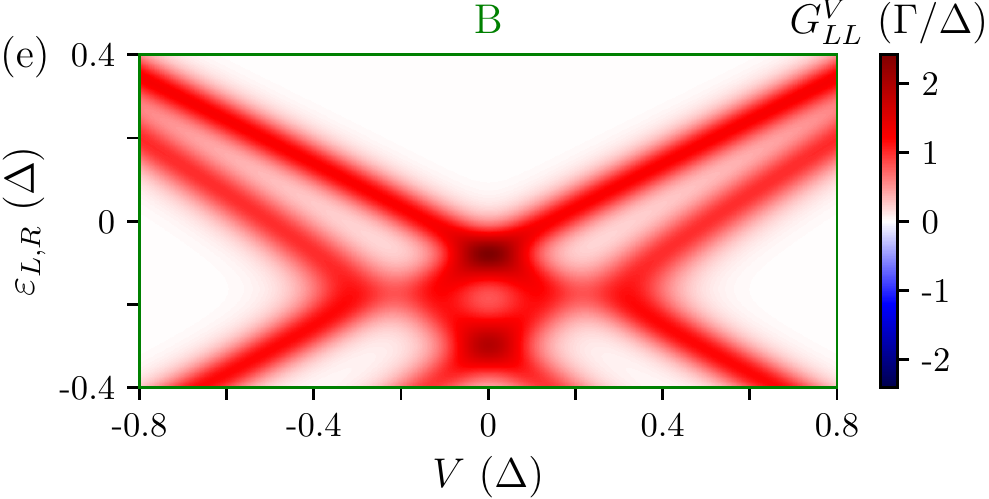}
	\caption{ (a), (b) and (c)~Local linear conductance $G_{LL}$ as a function of $\varepsilon_L$ and $\varepsilon_R$, with the same $\varepsilon_C$ as in Figs.~3(a, c, e) of the main text. (d) and (e)~ Local nonlinear conductance $G_{LL}^V$ as a function of $\varepsilon_L = \varepsilon_R$ and bias voltage $V$ with the same $\varepsilon_C$ as in (a) [(d)] and (b) [(e)].}
	\label{s_lcond}
\end{figure*}

\begin{figure}[h!]
    \centering
	\includegraphics[width=0.5\textwidth,trim={0cm 0cm 0.00cm 0.cm},clip]{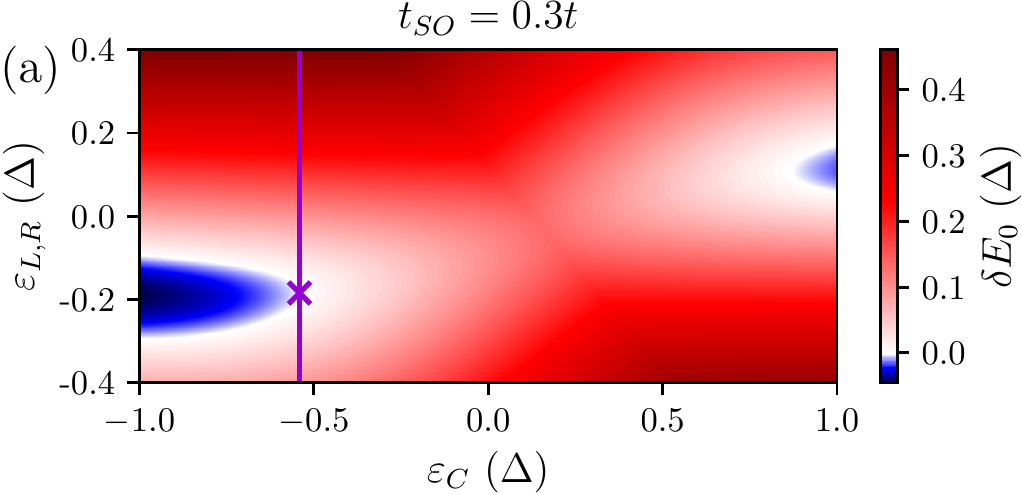}
	\includegraphics[width=0.5\textwidth,trim={0.0cm 0cm -0.0cm 0cm},clip]{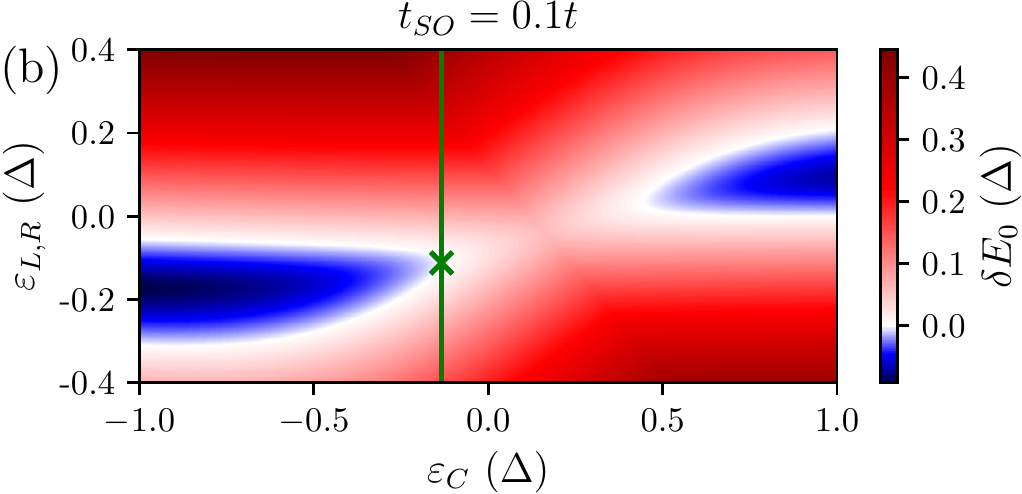}
	\includegraphics[width=0.21\textwidth,trim={0cm 0cm 0cm 0cm},clip]{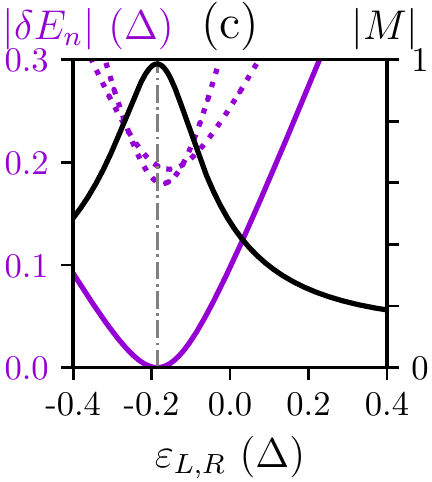}
	\includegraphics[width=0.21\textwidth,trim={0cm 0cm 0cm -0.cm},clip]{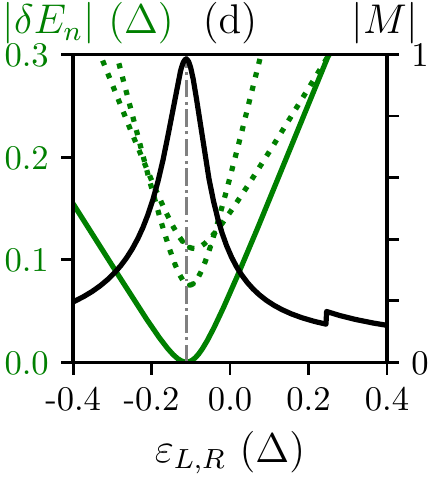}
	\caption{ (a) and (b)~$\delta E_0$ as a function of $\varepsilon_L = \varepsilon_R$ and $\varepsilon_C$ for $t_{SO} =0.3t $ and $t_{SO} =0.1t $ respectively. The cuts [purple line for (a) and green line for (b)] are  explored in (c) and (d). The purple cross in (a) marks the sweet spot at $\varepsilon_C \approx -0.541 \Delta$, $\varepsilon_L = \varepsilon_R \approx -0.185 \Delta$, while the  sweet spot in (b) is marked by the green cross at $\varepsilon_C \approx -0.134 \Delta$, $\varepsilon_L = \varepsilon_R \approx -0.112 \Delta$. (c)~$|\delta E_n|$ (left axis, purple lines) as a function of $\varepsilon_L = \varepsilon_R$ along the cut in (a) (purple line). The lowest excited state (full line) has different parity (odd in this case) than the ground state (even in this case); one of the two higher excited states shown (dotted lines) has even parity and the other odd. The black line shows $|M|$ (right axis) as a function of $\varepsilon_L = \varepsilon_R$ along the cut in (a). (d)~Same as (c) but along the cut in (b) (green line). The vertical dash-dotted lines in (c) and (d) indicate the maximum of $|M|$.}
	\label{s_tso}
\end{figure}

\begin{figure}[h!]
	\includegraphics[width=0.5\textwidth,trim={0cm -0.3cm 0.00cm 0.cm},clip]{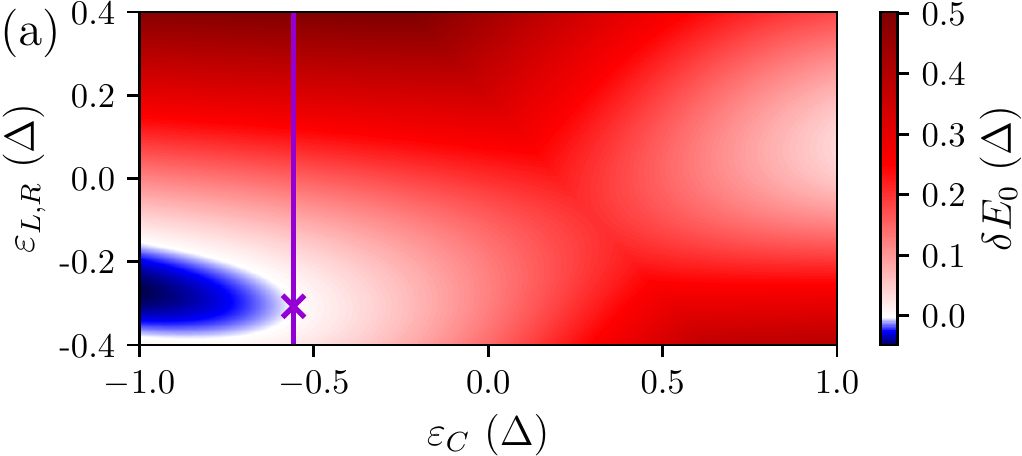}
	\includegraphics[width=0.20\textwidth,trim={0.0cm 0cm -0.0cm 0cm},clip]{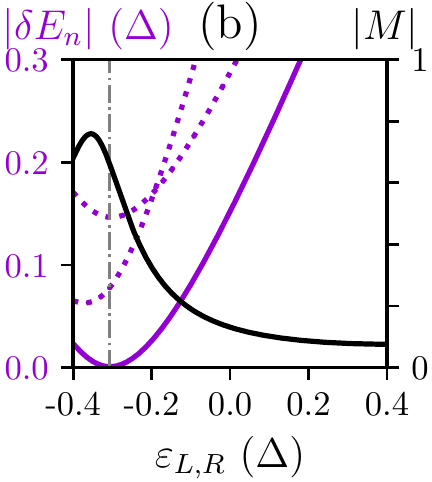}
	\includegraphics[width=0.26\textwidth,trim={0cm 0cm 0cm 0cm},clip]{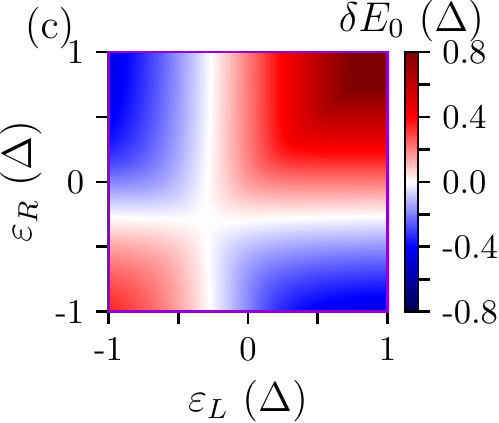}
	\caption{ (a)~$\delta E_0$ as a function of $\varepsilon_L = \varepsilon_R$ and $\varepsilon_C$ for $E_Z = 0.15 \Delta$ and $U = 5 \Delta$. The cut (purple line) is explored in (b). The purple cross marks the low-MP sweet spot at $\varepsilon_C \approx -0.558 \Delta$, $\varepsilon_L = \varepsilon_R \approx -0.316 \Delta$. (b)~$|\delta E_n|$ (left axis, purple lines) as a function of $\varepsilon_L = \varepsilon_R$ along the cut in (a) (purple line). The lowest excited state (full line) has different parity (odd in this case) than the ground state (even in this case); one of the two higher excited states shown (dotted lines) has even parity and the other odd. The black line shows $|M|$ (right axis) as a function of $\varepsilon_L = \varepsilon_R$ along the cut in (a). The vertical dash-dotted line in (b) indicates the misalignment between the points of maximum $|M|$ and of ground state degeneracy. (c)~Energy difference $\delta E_0$ between even and odd parity ground states as a function of $\varepsilon_L$ and $\varepsilon_R$, with the same $\varepsilon_C$ as in the purple line in (a).}
	\label{s_badsweet}
\end{figure}

\begin{figure}[h!]
    \centering
	\includegraphics[width=0.5\textwidth,trim={0cm 0cm 0.00cm 0.cm},clip]{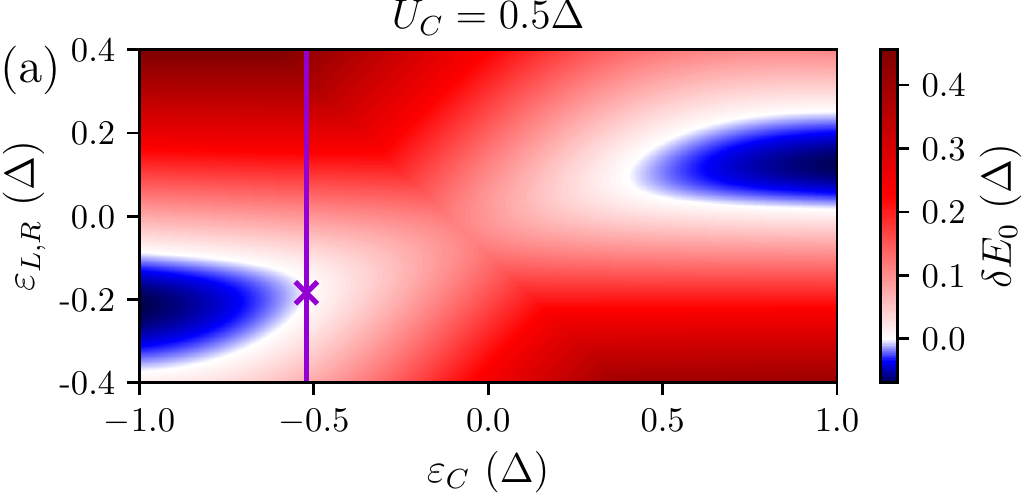}
	\includegraphics[width=0.5\textwidth,trim={0.0cm 0cm -0.0cm 0cm},clip]{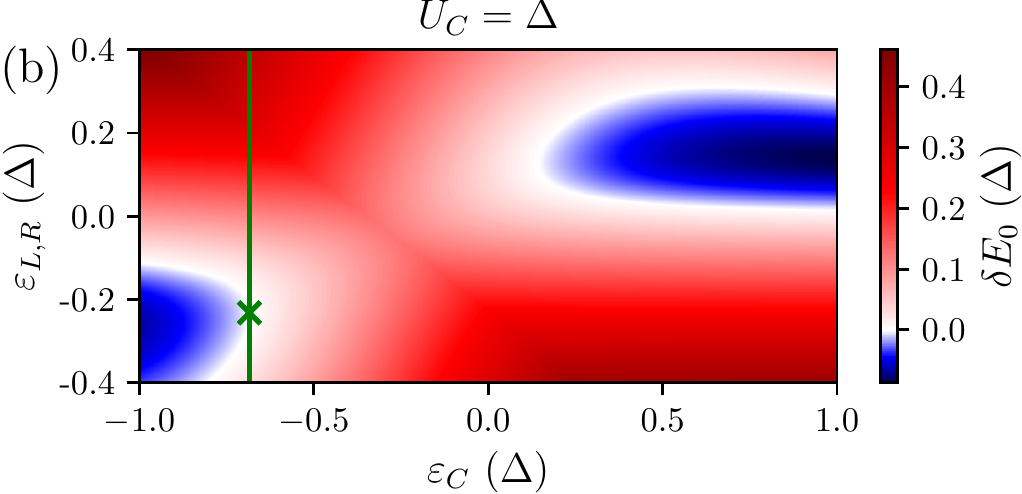}
	\includegraphics[width=0.21\textwidth,trim={0cm 0cm 0cm 0cm},clip]{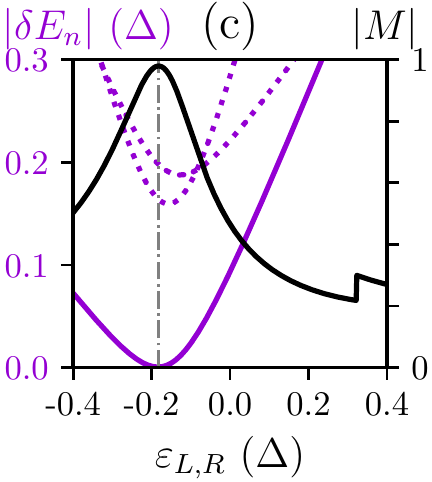}
	\includegraphics[width=0.21\textwidth,trim={0cm 0cm 0cm -0.cm},clip]{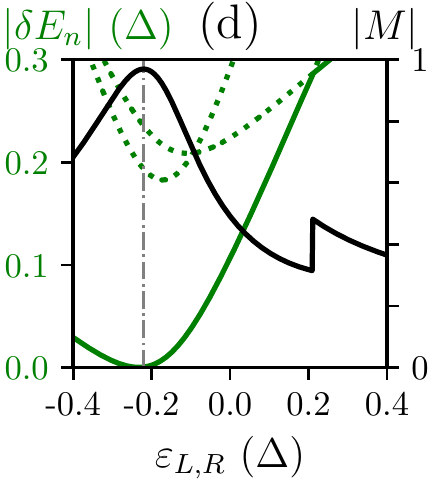}
	\caption{ (a) and (b)~$\delta E_0$ as a function of $\varepsilon_L = \varepsilon_R$ and $\varepsilon_C$ for $U_C = 0.5 \Delta $ in (a) and $U_C = \Delta $ in (b) ($E_{ZC}=0$). The cuts [purple line for (a) and green line for (b)] are  explored in (c) and (d). The purple cross in (a) marks the sweet spot at $\varepsilon_C \approx -0.522 \Delta$, $\varepsilon_L = \varepsilon_R \approx -0.184 \Delta$, while the  sweet spot in (b) is marked by the green cross at $\varepsilon_C \approx -0.685 \Delta$, $\varepsilon_L = \varepsilon_R \approx -0.232 \Delta$. (c)~$|\delta E_n|$ (left axis, purple lines) as a function of $\varepsilon_L = \varepsilon_R$ along the cut in (a) (purple line). The lowest excited state (full line) has different parity (odd in this case) than the ground state (even in this case); one of the two higher excited states shown (dotted lines) has even parity and the other odd. The black line shows $|M|$ (right axis) as a function of $\varepsilon_L = \varepsilon_R$ along the cut in (a). (d)~Same as (c) but along the cut in (b) (green line). The vertical dash-dotted lines in (c) and (d) indicate the maximum of $|M|$.}
	\label{s_Uc}
\end{figure}

\begin{figure}[h!]
    \centering
	\includegraphics[width=0.5\textwidth,trim={0cm 0cm 0.00cm 0.cm},clip]{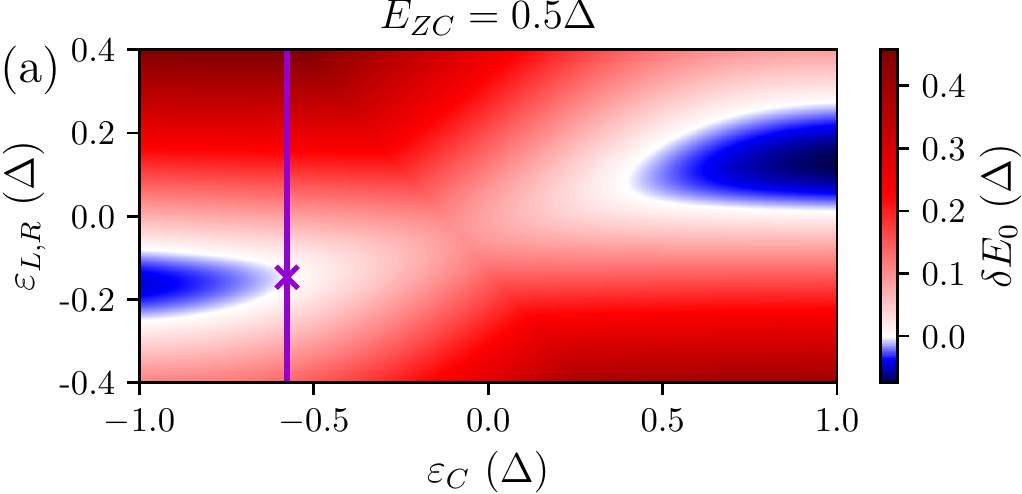}
	\includegraphics[width=0.5\textwidth,trim={0.0cm 0cm -0.0cm 0cm},clip]{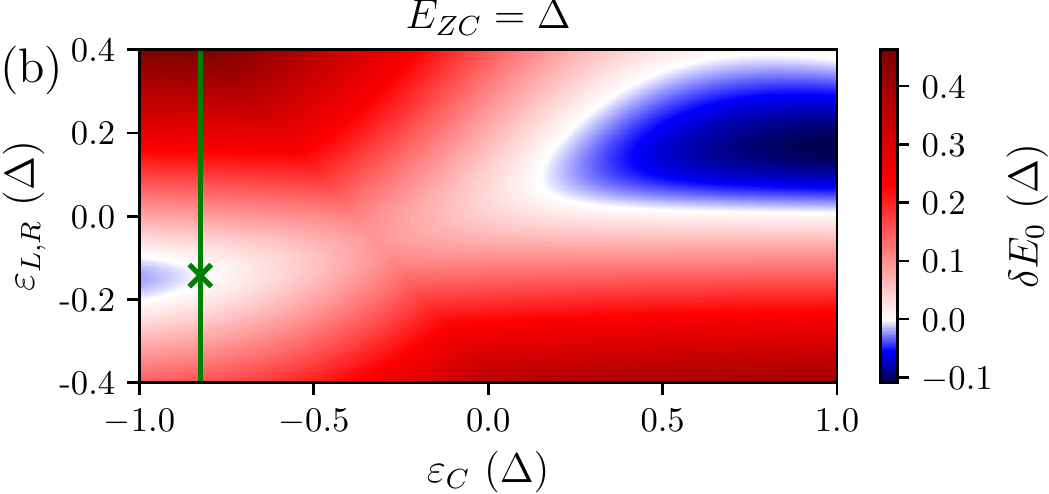}
	\includegraphics[width=0.21\textwidth,trim={0cm 0cm 0cm 0cm},clip]{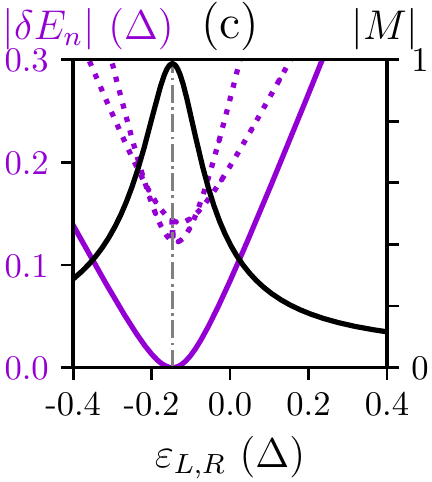}
	\includegraphics[width=0.21\textwidth,trim={0cm 0cm 0cm -0.cm},clip]{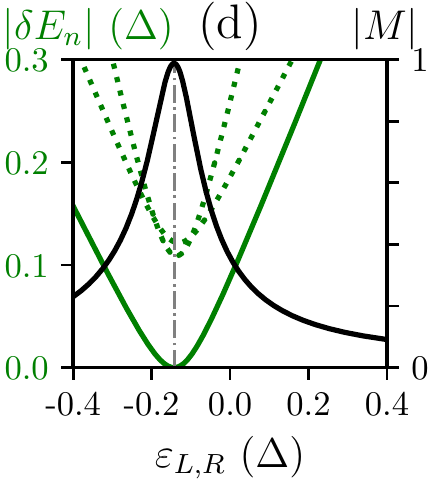}
	\caption{(a) and (b)~$\delta E_0$ as a function of $\varepsilon_L = \varepsilon_R$ and $\varepsilon_C$ for $E_{ZC} = 0.5 \Delta $ in (a) and $E_{ZC} = \Delta $ in (b) ($U_C=0$). The cuts [purple line for (a) and green line for (b)] are  explored in (c) and (d). The purple cross in (a) marks the sweet spot at $\varepsilon_C \approx -0.577 \Delta$, $\varepsilon_L = \varepsilon_R \approx -0.147 \Delta$, while the  sweet spot in (b) is marked by the green cross at $\varepsilon_C \approx -0.826 \Delta$, $\varepsilon_L = \varepsilon_R \approx -0.143 \Delta$. (c)~$|\delta E_n|$ (left axis, purple lines) as a function of $\varepsilon_L = \varepsilon_R$ along the cut in (a) (purple line). The lowest excited state (full line) has different parity (odd in this case) than the ground state (even in this case); one of the two higher excited states shown (dotted lines) has even parity and the other odd. The black line shows $|M|$ (right axis) as a function of $\varepsilon_L = \varepsilon_R$ along the cut in (a). (d)~Same as (c) but along the cut in (b) (green line). The vertical dash-dotted lines in (c) and (d) indicate the maximum of $|M|$.}
	\label{s_EZc}
\end{figure}

{\it Additional conductance plots.}
In the transport setup in Fig.~1 in the main text, we apply voltages $V_r$ to the two normal leads $r=L,R$. Because of the grounded superconducting lead (entering $H_{QDs}$ through pairing terms on QD $C$), $I_L \neq - I_R$ in general, and we calculate the currents $I_r(V_L, V_R)$ and conductances 
\begin{equation}
G_{r r'}(V_L, V_R) = \frac{dI_r(V_L, V_R)}{dV_{r'}}.    
\end{equation}
In the main text, Figs.~3(b, d, f) and Figs.~4(b, c) show the linear nonlocal conductance
\begin{equation}
G_{LR} = G_{LR}(0,0), 
\end{equation}

\noindent where we have introduced the notation $G_{r r'}(0,0) = G_{r r'}$.

Here, we show in Figs.~\ref{s_lcond}(a, b, c) $G_{LL}$ as a function of $\varepsilon_L$ and $\varepsilon_R$ for $\varepsilon_C$ values fixed by the cuts A, B and C in Figs.~3(a, c, e) of the main text. $G_{LL}$ shows peaks along the even-odd degeneracy lines. The nonlinear local conductance $G_{LL}^V = G_{LL}(V/2, - V/2)$ as a function of $\varepsilon_L = \varepsilon_R$ and a bias voltage $V$ for $\varepsilon_C$ corresponding to (a) and (b) is depicted in (d) and (e). The finite $V$ conductance lines reflect the excitation energies shown in Figs. 2(c) and (d) of the main text.

{\it Sweet spots for different spin-orbit coupling strengths.}
A significant spin-orbit interaction is a necessary ingredient to realize MBSs in the proposed system, as it provides a spin-flipping mechanism for electrons hopping between the QDs. Without spin-orbit coupling, $H_{QDs}$ would be diagonal in spin and there would be no CAR involving the lowest spin states on QDs $L$ and $R$. It is thus important to investigate the effect of a varying strength of the spin-orbit coupling on the quality of the MBSs, both in terms of the MP and the gap to the excited states. Similarly to Fig.~2(a) in the main text, Figs.~\ref{s_tso}(a, b) show $\delta E_0 = E^O-E^E$, the energy difference between the odd and even parity ground states, as a function of $\varepsilon_C$ and $\varepsilon_L = \varepsilon_R$, but for a larger and smaller spin-orbit coupling strength [$t_{SO} =0.3t $ in (a) and $t_{SO} =0.1t $ in (b)]. The cuts (lines) passing through the sweet spots (crosses) are marked with purple and green in (a) and (b) respectively. Like in Fig.~2(c) in the main text, Figs.~\ref{s_tso}(c) and (d) show $|M|$ together with $|\delta E_0|$ and the excitation energies above the global ground state ($|\delta E_n|, n \geq 1$) as a function of $\varepsilon_L = \varepsilon_R$ corresponding to the values of $\varepsilon_C$ marked with the lines in (a) and (b) [purple and green]. The peaks of $|M|$ occur at the same $\varepsilon_{L,R}$ as the ground state degeneracies in both (c) and (d), which implies high-MP sweet spots [$|M|>0.985$ in both (c) and (d)]. There exists a striking difference in the excitation energies though. While in (c) the lowest excitation is at around $0.2 \Delta$ - which is higher than the one presented in the main text - in (d) the excitation gap is smalller than $0.1 \Delta$. The reason for this dependence is that the couplings within the even parity sector are limited by the spin-orbit coupling strength. At the sweet spot, $\varepsilon_C$ has been adjusted to make the couplings in the even and odd sectors similar. These couplings set the distance to the lowest excited states. A good separation to the excited states (typically $\delta E_1 \gg T$) is required for MBS detection, as well as for qubit-like operations and nonabelian and nonlocal protocols.
%$|MP| \approx 0.986$ and $0.985$

{\it Low-MP sweet spots.}
In this part we provide additional information on the low-MP sweet spot whose nonlocal conductance is explored in Fig.~4(b) of the main text. Figure~\ref{s_badsweet}(a) is similar to Fig.~2(a) of the main text but with $E_Z = 0.15 \Delta$. The low-MP sweet spot is marked with a purple cross and the excitation energies together with $|M|$ along the cut (purple line) are plotted in Fig.~\ref{s_badsweet}(b). From the latter figure, it is clear that the peak of $|M|$ is not as pronounced as in Fig.~2(c) of the main text. More importantly, the even-odd degeneracy point does not coincide with the $|M|$ peak and corresponds to an even smaller $|M| \approx 0.661$. Additionally, the excitation gap is smaller than $0.1 \Delta$. The above features are characteristic for low-quality MBSs. In Fig.~\ref{s_badsweet}(c) we plot the difference between the energies of the even and odd ground states ($\delta E_0 = E^0 - E^E$) as a function of $\varepsilon_L$ and $\varepsilon_R$. The degeneracy lines do cross but are not straight as in Fig.~3(a) in the main text. This means that the ground state degeneracy is not entirely  protected from small changes of $\varepsilon_{L,R}$. We note that these features would probably be hard to discern with local conductance measurements.

{\it Sweet spots for finite $U_C$, $E_{ZC}$.}
For the results presented in the main text, we have considered a complete quenching of the Coulomb interactions and Zeeman splitting in QD $C$ $(U_C, \, \textcolor{black}{E_{ZC}}=0)$,  due to the strong coupling to the superconductor. Here, we \textcolor{black}{separately} study the effect\textcolor{black}{s} of finite $U_C, \, \textcolor{black}{E_{ZC}}$ on the MBSs quality.

Figures~\ref{s_Uc}(a) and (b) show $\delta E_0$ as a function of $\varepsilon_C$ and $\varepsilon_L = \varepsilon_R$ for $U_C =0.5\Delta$ and $U_C =\Delta$, respectively \textcolor{black}{($E_{ZC}=0$)}. The cuts (lines) passing through the sweet spots (crosses) are marked with purple in (a) and green in (b). Figures~\ref{s_Uc}(c) and (d) show $|M|$ together with $|\delta E_0|$ and the excitation energies above the global ground state ($|\delta E_n|, n \geq 1$) as a function of $\varepsilon_L = \varepsilon_R$ corresponding to the values of $\varepsilon_C$ marked with the lines in (a) and (b) (purple and green). The peaks of $|M|$ occur at the same $\varepsilon_{L,R}$ as the ground state degeneracies in both (c) and (d), which implies high-MP sweet spots ($|M| \approx 0.98$ and $|M| \approx 0.967$). The $|M|$ peaks are not as pronounced as in Fig.~2(c) of the main text $(|M| \approx 0.986)$, but the gap to the excited states is in fact enhanced [$|\delta E_1| \approx 0.2 \Delta$ in (d)]. Finite (but small) Coulomb interactions in QD $C$ can thus even be advantageous, as they provide a better separation between the ground state and the excited states without significantly disturbing the MBSs localization, as quantified by the MP. We have also verified that the results remain qualitatively unchanged in the presence of interdot Coulomb interactions. 

{\color{black} Figures~\ref{s_EZc}(a) and (b) show $\delta E_0$ as a function of $\varepsilon_C$ and $\varepsilon_L = \varepsilon_R$ for $E_{ZC} =0.5\Delta$ and $E_{ZC} =\Delta$, respectively \textcolor{black}{($U_C=0$)}. The cuts (lines) passing through the sweet spots (crosses) are marked with purple in (a) and green in (b). Figures~\ref{s_EZc}(c) and (d) show $|M|$ together with $|\delta E_0|$ and the excitation energies above the global ground state ($|\delta E_n|, n \geq 1$) as a function of $\varepsilon_L = \varepsilon_R$ corresponding to the values of $\varepsilon_C$ marked with the lines in (a) and (b) (purple and green). The peaks of $|M|$ occur at the same $\varepsilon_{L,R}$ as the ground state degeneracies in both (c) and (d), which implies high-MP sweet spots ($|M| \approx 0.987$ and $|M| \approx 0.988$). The $|M|$ peaks are as pronounced as in Fig.~2(c) of the main text, but the gap to the excited states is slightly reduced.}

{\color{black} {\it Emergence of Yu-Shiba-Rusinov states.} 
We note that Yu-Shiba-Rusinov (YSR) states  \cite{YU1965,Shiba1968,Rusinov1969} can be induced by the couplings of QDs $L$ and $R$ to the proximitized QD $C$. This physics is included in our calculations because we diagonalize the Hamiltonian of the three coupled QDs exactly. However, because of the strong coupling between the QDs, all states are strongly mixed and we do not find it useful to attempt to identify certain states as YSR-like. It is important to emphasize, however, that YSR physics related to the couplings of QDs $L$ and $R$ to QD $C$ does not prevent the appearance of MBSs. 

Possible YSR physics related to the coupling of QD $C$ to the bulk superconductor is, on the other hand, not included in the model because we consider the infinite gap limit for the bulk superconductor. The results presented in the main text are obtained for $E_{ZC} = U_{C} = 0$, in which case the coupling between QD $C$ and the bulk superconductor does not lead to YSR states. They might, however, become important for large enough $E_{ZC}$ and $U_C$. Our statements above and in the main text that our results remain qualitatively unchanged by finite $E_{ZC}$, $U_C$ are only verified for parameters small enough to not induce YSR states on QD $C$. Our expectation is, however, that such YSR states would not prevent the appearance of MBSs, as long as long as they do not approach zero energy, but they might change the exact value of $\varepsilon_C$ where the sweet spot appears. What is required for MBSs is just that the states in QD $C$ that mediate the couplings between QDs $L$ and $R$ do so in a way that is different in the even and odd parity subspaces, and that this difference can be controlled (for example by a gate voltage).}

{\color{black} {\it Sweet spots for an asymmetric system.} 
In the main text we have considered a symmetric system, $t_L=t_R=t$,  $t^{SO}_L=t^{SO}_R=t_{SO}$, $U_L=U_R=U$ and we have only presented MP results for $\varepsilon_L = \varepsilon_R$. Introducing any asymmetry, the definition of MP remains valid but $|M_L| \neq |M_R|$. A different value of the MP in the two QDs should be interpreted as a different degree of localization of the two MBSs. The sweet spots are then found at points where $\varepsilon_L \neq \varepsilon_R$. $|M_L|$ and $|M_R|$ can still come very close to $1$ at those sweet spots, but they are not completely identical.}

% What I initally wrote
%{\color{blue} With a finite $U_C$ or $E_{ZC}$ unpaired spins in QD $C$ can lead to the formation of YSR states due to the stong coupling with the superconductor. In our model we consider the ``infinite superconducting gap'' limit and we thus cannot account for the formation of such states in QD $C$. YSR states can play a role for $U_C$, $E_{ZC} \neq 0$, but the general mechanism leading to the formation of MBSs would not be disrupted. On the other hand, YSR physics is included in the model for QDs $L$ and $R$ if one views QD $C$ as a superconducting lead in the zero bandwidth approximation (see for example Ref. \cite{Grove_NatCom2018}). Despite the fact that YSR states can appear, we find localized MBSs in the system.}

%\newpage
%\clearpage

\bibliography{supplemental_material}% Produces the bibliography via BibTeX.